\documentclass{article}

\usepackage{amsfonts}
\usepackage{epsfig}

\begin{document}


\title{\textsf{\textbf{Transitions to Intermittency and Collective Behavior\\ in
Randomly Coupled Map Networks}}}

\author{D. Volchenkov\footnote{dima.volchenkov@uni-bielefeld.de},
 S. Sequeira\footnote{sandra@mathematik.uni-bielefeld.de}, Ph. Blanchard, \\
\em Zentrum f\"{u}r interdisciplin\"{a}re Forschung (ZiF) \texttt{and}\\
\em  BiBoS  \texttt{and} FSPM - Strukturbildungsprozesse, \\ \em
University Bielefeld,  Postfach 100131, D-33501, Bielefeld, Germany \\and \\ M.G.
Cosenza \\ \em Centro de Astrofisica Te\'orica, Universidad de Los Andes,\\ \em A.
Postal 26 La Hechicera, Merida 5251, Venezuela}

\date{\today}
\maketitle

\begin{abstract}
We study the transition to spatio- temporal intermittency in
networks of randomly coupled Chat\'e-Manneville maps. The
relevant parameters are the network connectivity, coupling
strength, and the local parameter of the map. We show that
spatiotemporal intermittency occurs for some intervals or windows
of the values of these parameters. Within the intermittency
windows, the system exhibits periodic and other nontrivial
collective behaviors. The detailed behavior depends crucially
upon the topology of the random graph spanning the network. We
present a detailed analysis of the results based on the
thermodynamic formalism and random graph theory.

\end{abstract}
\vspace{0.5cm}
\large

\leftline{\textbf{Keywords:} \textsl{coupled chaotic  maps, random graphs,\\  phase
transitions, nontrivial collective}} \leftline{ \textsl{behavior, spatiotemporal
intermittency}}

\leftline{\textbf{ PACS codes: } \textsl{05.45.+b, 05.70.Fh.}}

\leftline{\textbf{ AMS codes:} \textsl{82C26, 58F11.}}

\vspace{0.5cm}

\vspace{0.5cm}

\pagebreak

\section{Introduction.}
\label{sec:intro}
\noindent

Partial differential equations describing continuous models and real physical systems
can, in many cases, be discretized into a system  of coupled map lattices (CML).
Coupled map lattices are spatiotemporal dynamical systems comprised of an interacting
array of discrete-time maps. Much attention to these systems has been drawn in virtue
of studies of generic properties of spatiotemporal chaos, \cite{1},\cite{2}. A
mean-field extension of CML is the globally coupled map lattice introduced by Kaneko
\cite{Kan90}. Here we consider another mean-field extension which refers to random
networks of coupled maps.

Although just a few studies devoted to randomly coupled chaotic map networks (RCMN)
have been reported by this time, it is beyond dispute that such systems would be very
rich in practical applications. To motivate the increasing interest RCMN, one has to
note that most real-world networks are of a disordered nature. Social networks
\cite{WF}, biological communities forming food webs \cite{food}, and, moreover,
computer networks \cite{WWW}, to name just few, have plenty of random shortcuts
inconsistent with any regular structure. In this context, interesting investigations
of coupled map systems defined on non-uniform lattices have been reported in \cite{CK}
(CMLs on a Sierpinski gasket) and in \cite{GC} (CMLs on a Cayley tree). In view of
this, an ensemble of maps coupled at random would provide a forthright model for
studying various properties of these disordered networks. The thorough investigation
of RCMN would shed new light on the problem of spatiotemporal behavior of discrete
extended systems having infinitely many degrees of freedom.

To our knowledge, randomly coupled logistic maps $f(x)=ax(1-x)$ have been considered
first in \cite{CM1}. The emergence of synchronization in random networks of logistic
maps with nonlocal couplings has been investigated in \cite{Gade}, and more recently,
dynamical clustering has been observed in maps connected symmetrically at random
\cite{MM}.

We study the collective behavior and phase transitions in a RCMN different from those
considered in \cite{CM1,MM}. The somewhat ``statistically simplest" RCMN is
considered. On one hand, the Chat\'{e}-Manneville map (CM) \cite{CM} which we use as a
local evolution law can be either in a chaotic or ``turbulent" (excited) state, or in
a fixed point or ``laminar" (inhibited) state. On the other hand, the network topology
in our model is spanned by a random graph $\mathbb{G}(N,k)$, corresponding to $N$
sites and such that each site has precisely $k$ outgoing edges.

In the present article, we show that the entire collective behavior is the net result
of the interplay between the properties of local map and the probabilistic topology of
relevant random graph. Let us note that, in the domain of coupled chaotic maps, the
notion of phase transition has been traditionally applied to at least two different
classes of phenomena. The first class constitutes the case when a valuable fraction of
nodes in the lattice becomes either excited or inhibited at some critical values of
the parameters. We shall call these situations either as a transition to intermittency
or to relaminarization. The second class refers to the appearance of global periodic
motion within a sustained turbulent state. We shall call it as a transition to
collective behavior.

In Sec.~\ref{sec:definition}, the random networks of coupled maps explored in this
article are introduced. Section~\ref{sec:results} presents the phenomena of
spatiotemporal intermittency and nontrivial collective behavior found by direct
simulations on randomly coupled Chat\'e-Manneville maps. In Secs~\ref{sec:geom},
\ref{sec:TDformRCMN}, and \ref{sec:transitions}, the observed behavior is analyzed
through a theoretical framework. Our approach is twofold. First, we develop a
thermodynamic formalism (TD) for RCMN. Secondly, we use the random graph theory
invented by P. Erd\"{o}s and A. R\'{e}nyi \cite{rgstart}, and which has become a basis
for discrete mathematics located at the intersection of graph theory, combinatorics,
and probability theory \cite{Ballobas}, \cite{RGBook}. Finally,
Sec.~\ref{sec:conclusion} contains the conclusions of this work.

\section{Coupled maps on random networks.}
\label{sec:definition} \noindent


Let $\Omega\subset\mathbb{Z}$ be the finite lattice of $N\in \mathbb{N}$ sites. At
each site $\omega\in\Omega$ there is a local phase space $\mathrm{X}_{\omega}$ with an
uncountable number of elements. The global phase space
$\cal{M}=\Pi_{\omega\in\Omega}\mathrm{X}_{\omega}$ is a direct product of local phase
spaces such that a point $\mathrm{x}\in \cal{M}$ can be represented as
$\mathrm{x}=(x_{\omega})$. A \textsl{coupled map lattice} is any mapping
$\Phi:\mathrm{M}\to \mathrm{M}$ which preserves the product structure, $\Phi
\mathrm{x}=(\Phi_{\omega}x)_{\omega\in \Omega},$ in which $\Phi_{\omega}:\cal{M}\to
\mathrm{X}_{\omega}.$ The mapping, $\Phi=G\circ F,$ is a composition of an independent
local mapping $(Fx)_\omega=f_\omega(x_\omega),$ $f_\omega:\mathrm{X}_\omega\to
\mathrm{X}_\omega,$ and an interaction, $(Gx)_\omega=g_\omega(x)$.

We consider the following coupled map lattice supplied with some boundary conditions,
\begin{equation}\label{cml}
(\Phi x)_\omega=
\left[(1-\varepsilon)\mathbf{I}+\frac{\varepsilon}{k}\mathbf{M}\right]f(x_\omega),
\end{equation}
where $\varepsilon\in [0,1]$ is the coupling strength parameter,
 $0< k< N-1$ is the connectivity number,
$\mathbf{I}$ is a unit matrix, and $\mathbf{M}$  is a traceless
connectivity matrix, $\mathrm{M}_{jj}=0,$ determining the network
topology.


Some models of coupled maps on different random network architectures have been
proposed in the literature \cite{CM1,Gade,MM}. Let us note that because of the
casuality property, coupled map systems are related to \textsl{ directed} random
graphs. In \cite{CM1}, the connectivity $\mathcal{K}$ is kept fixed, and the
connectivity matrix $M_{i,j}$ is not necessary symmetric (if $j$ is a neighbor of $i$,
the reverse may not be true). This random network refers to a \textsl{ uniform
directed random graph}, denoted by $\mathbb{G}(N,\mathcal{K})$ defined on the vertex
set $[N]$ with exactly $\mathcal{K}$ edges. Denoting the family of all such graphs as
$\mathcal{G},$ we obtain a uniform probability distribution to observe a particular
realization $\mathbb{G}(N,\mathcal{K}),$ $$
\mathbb{P}(\mathbb{G})=\left(\begin{array}{c}\left(
  \begin{array}{c}
    N \\
    2 \
  \end{array}\right)\\
  \mathcal{K}
\end{array}
\right)^{-1}, \quad \mathbb{G}\in \mathcal{G}$$.

Two models have been considered in \cite{CM1}. In the first one, there is a "frozen
disorder" with a fixed graph topology configuration. In the second model, new
connections are drawn at each time step. From the random graph theory, the second
model is known as a random graph \textsl{process}
$\left\{\mathbb{G}(N,\mathcal{K})\right\}_{\mathcal{K}}$ which begins at time $0$ and
adds new edges, one at a time. The $\mathcal{K}$-th stage of this Markov process can
be identified with the uniform random graph $\mathbb{G}(N,\mathcal{K})$ as it evolves
with $\mathcal{K}$ growing from $0$ to $\left(\begin{array}{c}
  N \\
  2
\end{array}\right)$.

In \cite{MM}, the random connectivity matrix is symmetric ($M_{ij}=M_{ij}$), and the
matrix elements are either $0$ (when the connection between maps $i$ and $j$ is
absent) or $1$ (if otherwise), while loops are not allowed, ($M_{ii}=0$). The main
advantage of this model is the independent presence of edges, but the drawback is that
the number of edges is not fixed, but varies according to a binomial distribution with
an expectation $ \left(\begin{array}{c}
  N \\
  2
\end{array}\right)p$.
This model relies upon a binomial random directed simple symmetric graph,
$\mathbb{G}(N,p)$, $0\leq p\leq 1$.

Another model of a random matrix has been studied in \cite{Gade}. In this case, the
connectivity matrix element $M_{ij}$ is equal to the number of times map $i$ is
connected to map $j$, i.e.,  possible multiple edges and loops have been taken into
account. Therefore, $M_{ij}$ is not necessarily symmetric, and $\sum _iM_{ij}=k$ for
any $j$, i.e. each map is coupled to $k$ maps chosen randomly (it can be coupled to
itself). We denote such a random directed graph as $\mathbb{G}^*(N,k)$.

\begin{figure}[ht]
 \noindent
 \begin{minipage}[b]{.36\linewidth}
 \begin{center}
 \epsfig{file=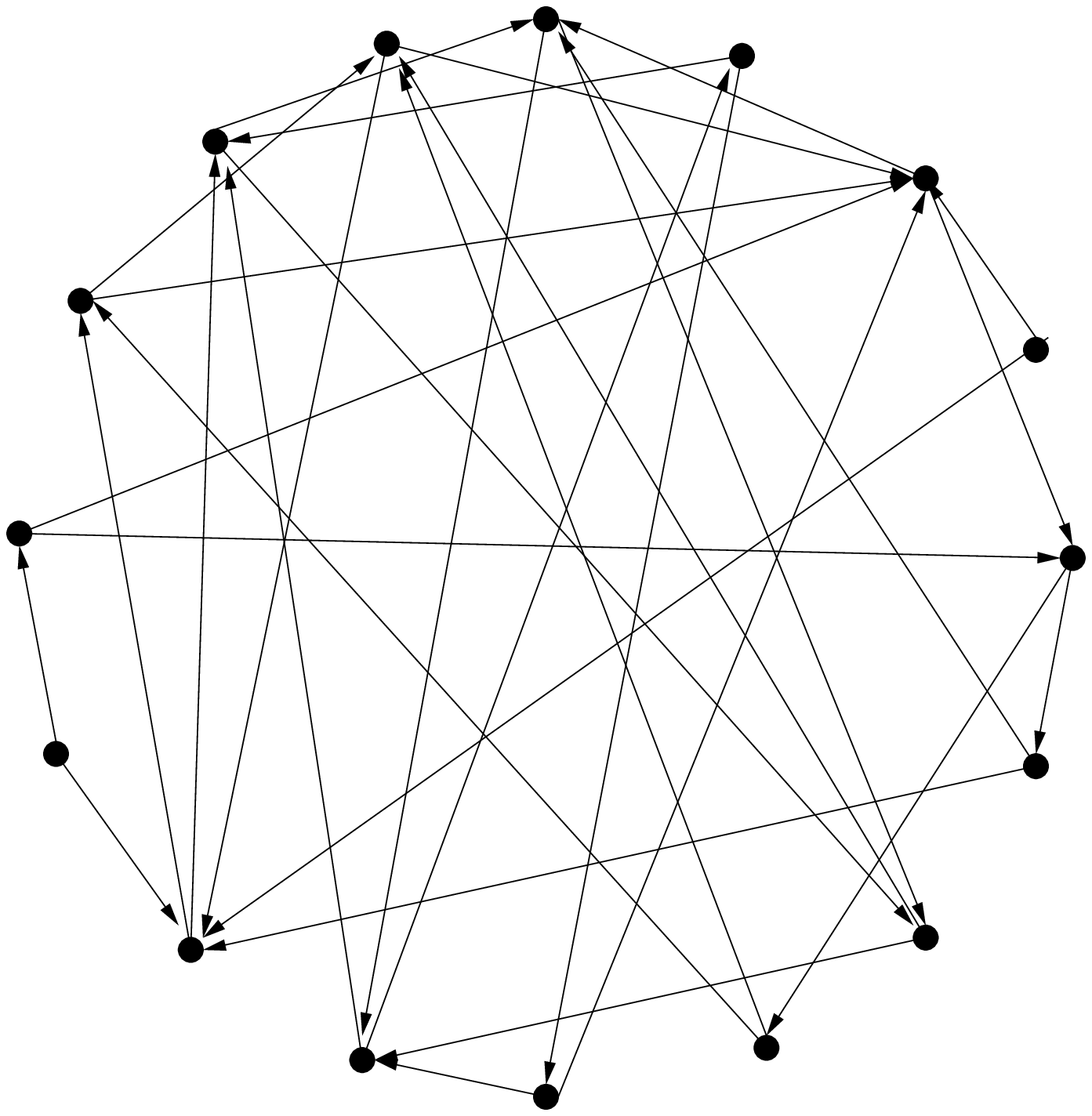, angle= 0}
 \end{center}
\end{minipage}
\caption{A realization of the random graph $\mathbb{G}(16,2).$} \label{F4}
\end{figure}

In the present paper, we consider a scheme such that the  elements of the connectivity
matrix are taken to be either $0$ or $1$, and the diagonal elements are always taken
as $0$, i.e. the coupling to itself is ruled out. The number of units in each row of
the conectivity matrix is fixed at $k\in [1,N-1]$. Each vertex $\omega$ in the
relevant random graph has always $k$ outgoing edges. The number of incoming edges is a
random Poisson distributed variable with a mean $z=kN/(N-1).$ We denote such a random
directed graph as $\mathbb{G}(N,k)$. A random realization of $\mathbb{G}(16,2)$ is
given in Fig.~(\ref{F4}).

Random graphs $\mathbb{G}(N,1)$ have been extensively studied in
\cite{KolchinCo}-\cite{AP}. However, many properties of $\mathbb{G}(N,k)$ for
arbitrary $k$ remain to be investigated. A convenient property of such graphs is that
they allow an explicit computation of the graph entropy \cite{Codbook} as
$h\left(\mathbb{G}(N,k)\right)=\log_2 k$.

Let us note that in the limit $N\to \infty$, the graph $\mathbb{G}(N,k)$ is
asymptotically equivalent to $\mathbb{G}^*(N,k)$ considered in \cite{Gade} since
either possibility, that two sites will be connected more than once or that one site
will be coupled to itself, are negligible. If $\left(\begin{array}{c}
  N \\
  2
\end{array}\right)p \approx k$, the graph $\mathbb{G}(N,k)$ is also asymptotically
equivalent to $\mathbb{G}(N,p)$ (i.e., to a binomial random
directed graph). However, it differs substantially from
$\mathbb{G}(N,p)$ considered in \cite{MM} since we have
$M_{ij}\ne M_{ji}$. The properties of $\mathbb{G}(N,k),$ in
general,  turn out to be quite different from those of either
binomial random graphs or uniform random graphs (i.e. having the
total number of edges fixed). For example, $\mathbb{G}(N,k)$ are
typically sparse but connected.

\section{Spatiotemporal intermittency and collective behavior.}
\label{sec:results}
\noindent

Spatiotemporal intermittency in extended systems consists of a sustained regime where
coherent and chaotic domains coexist and evolve in space and time. The transition to
turbulence via spatiotemporal intermittency has been studied in coupled map lattices
whose spatial supports are Euclidean \cite{CM1,Chate1,Kan,Stassi}, and also in
nonuniform lattices such as fractals \cite{CK2} and hierarchical lattices \cite{CT}. A
local map possessing the minimal requirements for observing spatiotemporal
intermittency is the Chat\'e-Manneville map \cite{CM1}
\begin{equation}
\label{map} f(x)=\left\{
\begin{array}{ll}
\frac{r}{2}\left(1-\left| 1-2x \right|\right), & \mbox{if $x \in [0,1]$} \\ x, &
\mbox{if $x > 1$},
\end{array}
\right.
\end{equation}
with $r>2$. This map is chaotic for $f(x)$  in $[0,1]$. However, for $f(x) >1$ the
iteration is locked on a fixed point. The local state can thus be seen as a continuum
of stable ``laminar" fixed points $(x>1)$ adjacent to a chaotic repeller or
``turbulent" state $(x \in [0,1])$.

In regular arrays, the turbulent state can propagate through the lattice in time for a
large enough coupling, producing sustained regimes of spatiotemporal intermittency
\cite{CM1,Chate1}. Here, we investigate the phenomenon of transition to turbulence in
random networks $\mathbb{G}(N,k)$ using the local map $f$ (Eq.(\ref{map})) in the
coupled system described by Eq~(\ref{cml}). As observed for regular lattices, starting
from random initial conditions and after some transient regime, our systems settle in
a stationary statistical behavior. The transition to turbulence can be characterized
through the average value of the instantaneous fraction of turbulent sites $F_{t}$, a
quantity that serves as the order parameter \cite{CM1}. We have calculated $\langle F
\rangle$ as a function of the coupling parameter $\epsilon$ for several random
networks from a time average of the instantaneous turbulent fraction $F_t$, as
\begin{equation}
\label{<<F>>}
\langle F \rangle={1 \over T} \sum_{t=1}^T F_t.
\end{equation}
About $10^4$ iterations were discarded before taking the time average in
Eq.~(\ref{<<F>>}), and $T$ was typically taken at the value $10^4$.

We consider Chat\'e-Manneville maps coupled on a random network $\mathbb{G}(N,k)$ for
different parameter values. As initial conditions, we use random cell values uniformly
distributed over the interval $[0,r/2]$. Some minimum number of initially excited
cells is always required to reach the sustained turbulent state. The typical system
size used in the calculations was $N=10^4$. We have verified that increasing the
averaging time $T$ or the network size $N$ do not have appreciable effects on the
results.

Two models of random topological configuration have been studied. Model A purposes a
random graph to be fixed while the maps are updating. It is, in fact, equivalent to a
model of "frozen disorder" proposed in \cite{CM1}. Model B possesses a random graph
which is changed at each time step simultaneously with the updating of the maps.

\begin{figure}[ht]
 \noindent
 \begin{minipage}[b]{.36\linewidth}
 \begin{center}
 \epsfig{file=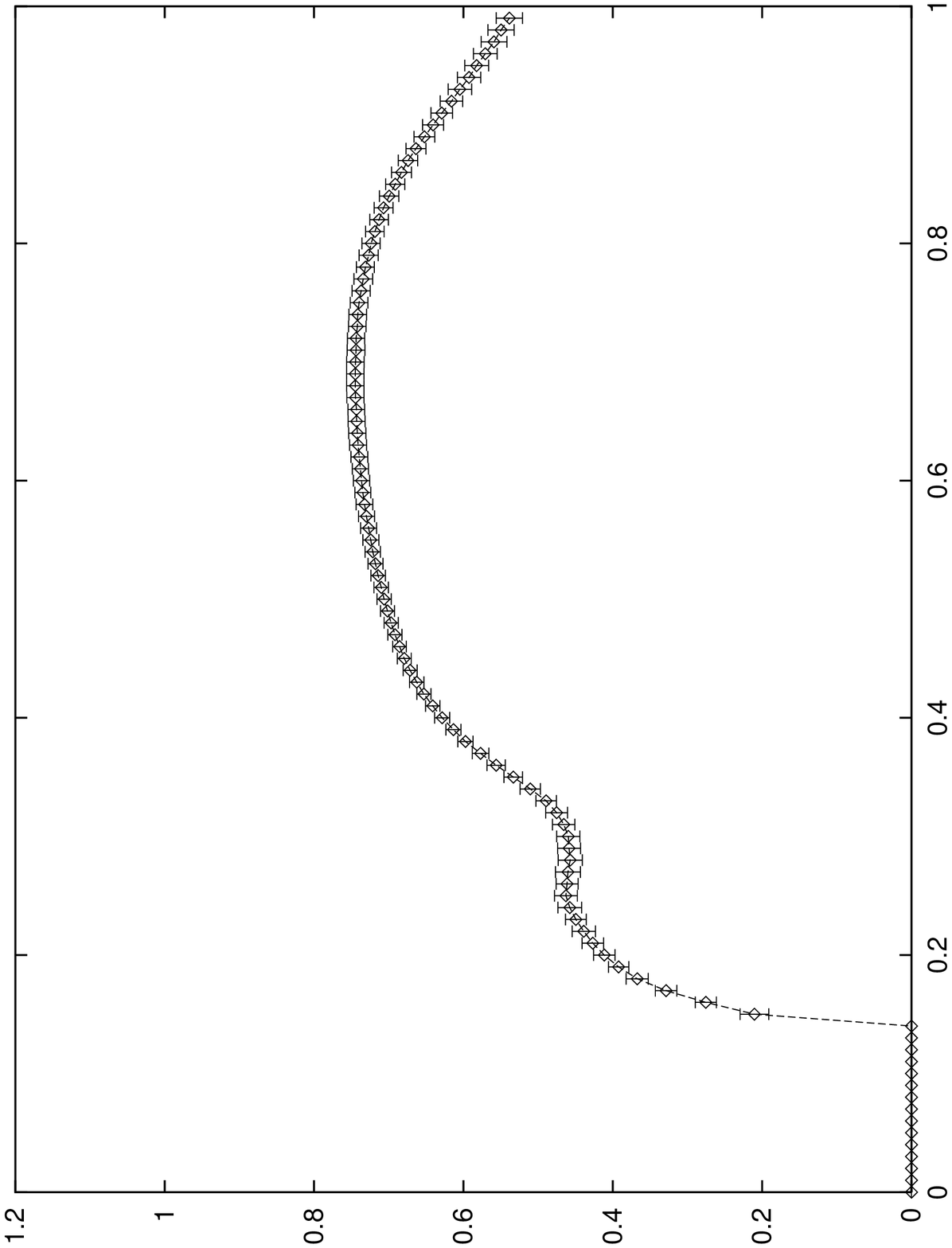, angle= 0}
 \end{center}
\end{minipage}
\caption{The mean turbulent fraction $<F>$ vs. $\varepsilon$, for $r=3$. Model A,
$\mathbb{G}(10^4,2)$. Onset of spatiotemporal intermittency.} \label{F7}
\end{figure}

We have calculated $\langle F \rangle $ vs. $\epsilon$ for random networks with
different connection numbers $k$. The local parameter has been kept fixed at $r=3$ in
most of the calculations. Figure~(\ref{F7}) shows the mean turbulent fraction $\langle
F\rangle$ versus $\varepsilon$ for $\mathbb{G}(10^4,2)$. One can see that, as
$\varepsilon>\varepsilon_c\approx 0.145$, the excitation occupies a significant
fraction of vertices. The random graph $\mathbb{G}(10^4,2)$ consists of a set of small
disjoint subgraphs of the length ($m\ll N$) and the largest connected component which
includes about $O(N^{2/3})$ vertices \cite{RGBook}. The transition to spatiotemporal
intermittency for $k=2$ is characterized by the scaling relation $\langle
F\rangle\propto (\varepsilon-\varepsilon_c)^\beta$ near the critical value
$\epsilon_c$, where the critical exponent is $\beta = 0.55\pm 0.03$ for $r=3$.

A power law behavior of mean turbulent fraction near the onset of spatiotemporal
intermittency also occurs for diffusively coupled CM maps in regular Euclidean
lattices (i.e., nearest neighbor coupling) \cite{CM,Chate1}. The value of the critical
exponent $\beta$ for the random network with $k=2$ coincides with that found for the
two-dimensional lattice \cite{Chate1,HWJ}.

\begin{figure}[ht]
 \noindent
 \begin{minipage}[b]{.36\linewidth}
 \begin{center}
 \epsfig{file=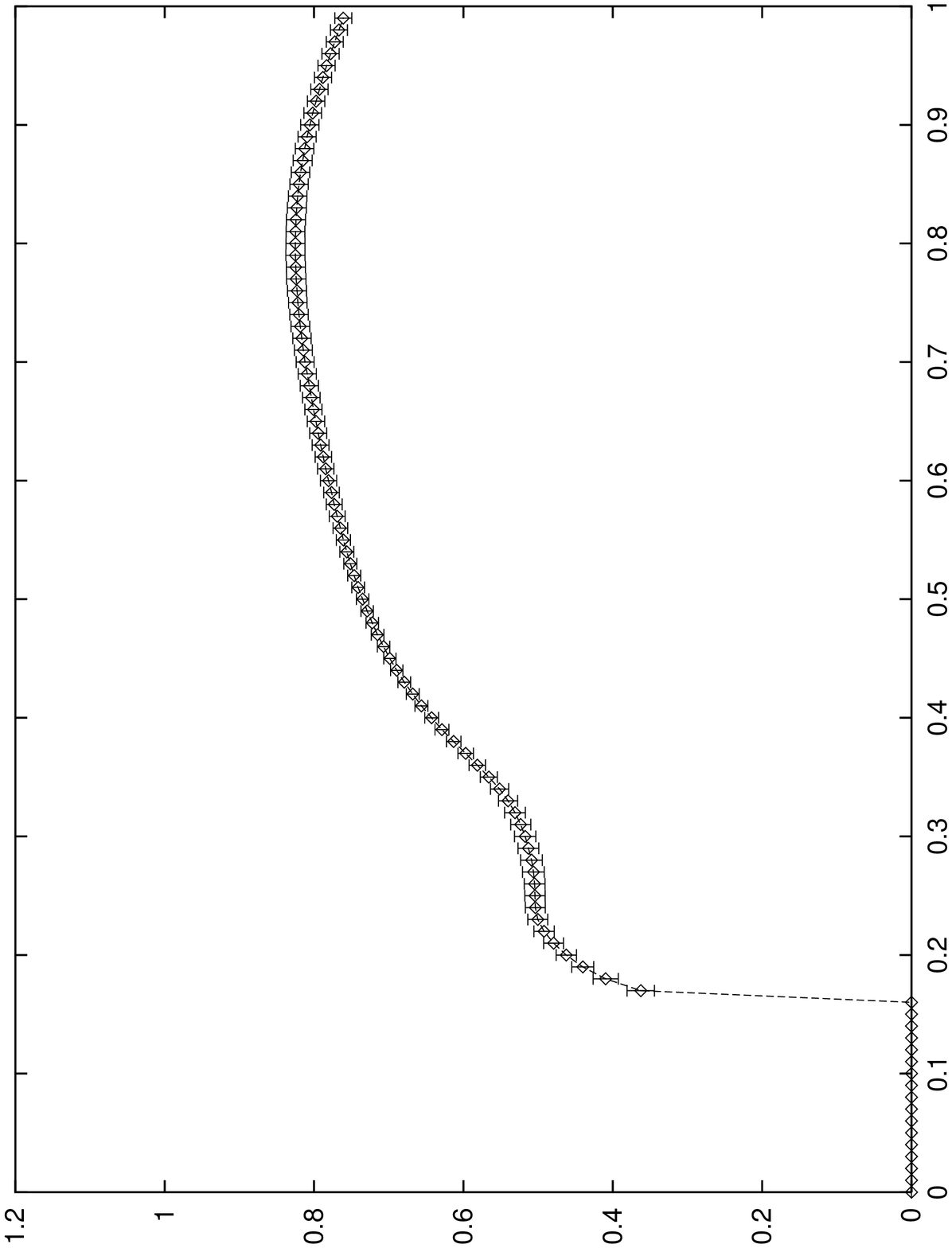, angle= 0}
 \end{center}
\end{minipage}
\caption{Model A, $\mathbb{G}(10^4,3)$ . The mean fraction $<F>$ vs. $\varepsilon$,
for $r=3$. The critical coupling for the onset of intermittency is
$\varepsilon_c\approx 0.161$.} \label{F8}
\end{figure}

For $k=3$, a Hamilton cycle traversing all vertices in the network appears for the
first time. There is no isolated vertex in the graph $\mathbb{G}(10^4,3)$.
Fig.~(\ref{F8}) shows that the onset of intermittency for the case $k=3$ occurs more
abruptly as $k$ is increased.

\begin{figure}[ht]
 \noindent
 \begin{minipage}[b]{.36\linewidth}
 \begin{center}
 \epsfig{file=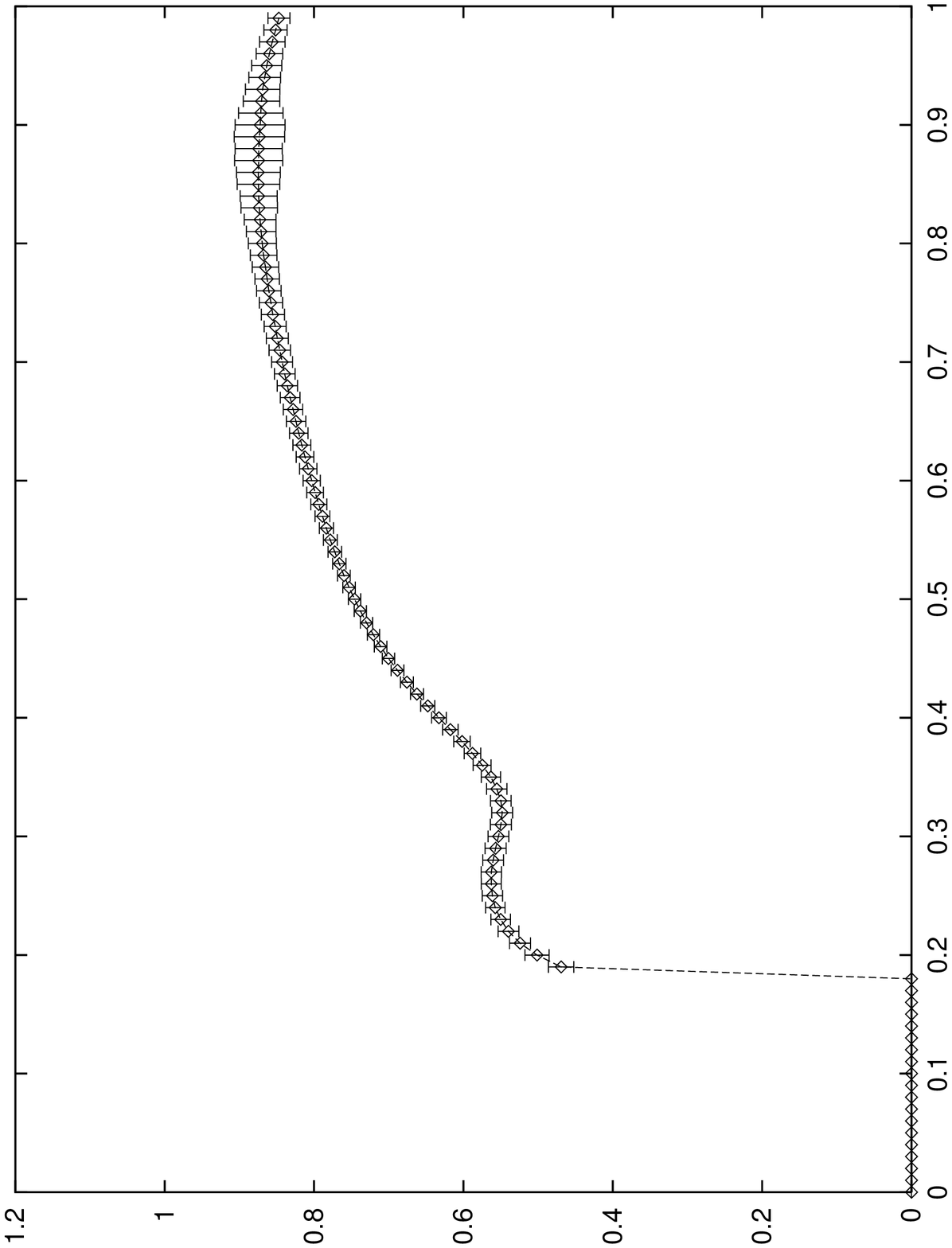, angle= 0}
 \end{center}
\end{minipage}
\caption{The average turbulent fraction $<F>$ vs $\varepsilon$ in the network
$\mathbb{G}(10^4,4)$, Model A, with $r=3$. The onset of nontrivial collective behavior
is observed as an emerging ``bulb" around $\varepsilon=0.85$.} \label{F9}
\end{figure}
\begin{figure}[ht]
 \noindent
 \begin{minipage}[b]{.36\linewidth}
 \begin{center}
 \epsfig{file=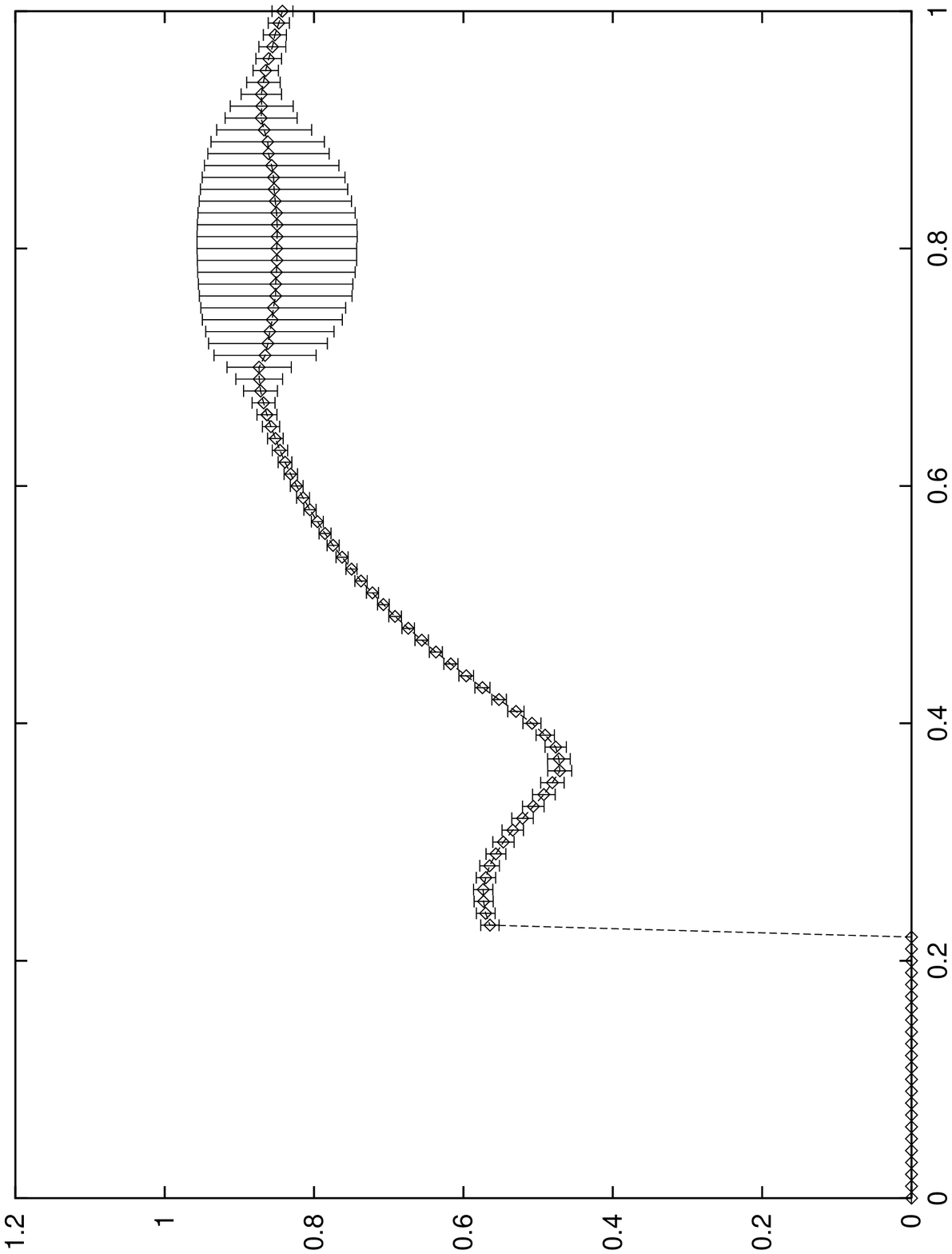, angle= 0}
 \end{center}
\end{minipage}
\caption{The average turbulent fraction $<F>$ vs. $\varepsilon$ with fixed $r=3$ for
$\mathbb{G}(10^4,4)$, Model B. Nontrivial collective behavior occurs in ``bulb"
region.} \label{F11}
\end{figure}

Figures~(\ref{F9}) and (\ref{F11}) display the mean turbulent fraction $\langle
F\rangle$ versus the coupling $\varepsilon$ for both Model A and Model B in the RCMN
induced by realizations of the random graph $\mathbb{G}(10^4,4)$. Figs.~(\ref{F9}) and
(\ref{F11}) show that the onset of intermittency when $k=4$ occurs as a discontinuous
jump in the order parameter $\langle F \rangle$ at the critical value of the coupling.
A discontinuous jump of $\langle F\rangle$ at the onset of spatiotemporal
intermittency has also been observed for globally coupled Chat\'e-Manneville maps and
interpreted as a first order phase transition in \cite{CP}.

The error bars shown on $\langle F\rangle$ in Figs.~(\ref{F9}) and (\ref{F11})
correspond to the standard deviation (the square root of the variance) of the time
series of the instantaneous fraction $F_t$ at each value of $\varepsilon$. With
increasing system size $N$, some of those fluctuations do not fade out. Large,
non-statistical fluctuations in the time series of the instantaneous turbulent
fraction $F_t$ persist with increasing connectivity $k$ in the networks. For
$\varepsilon>0.5$ these fluctuations appear as large ``bulbs" around $\langle
F\rangle\approx 1$. This phenomenon is associated to the emergence of nontrivial
collective behavior commonly observed in CML systems \cite{Chate2}. In fact, the
observed large amplitudes of the standard deviations reflect collective periodic
states of the system.

\begin{figure}[ht]
 \noindent
 \begin{minipage}[b]{.36\linewidth}
 \begin{center}
 \epsfig{file=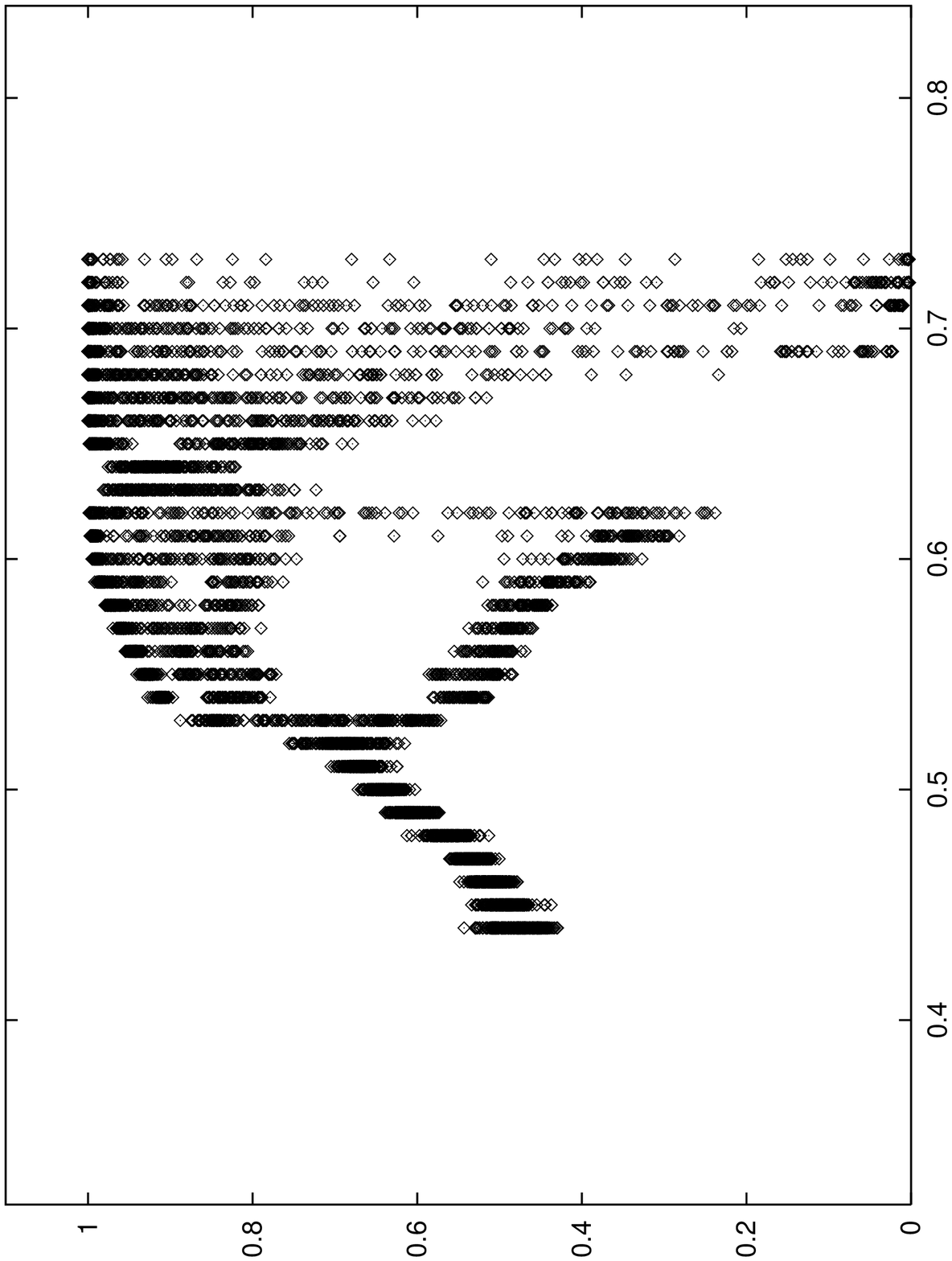, angle= 0}
 \end{center}
\end{minipage}
\caption{ Model A. $\mathbb{G}(10^4,25)$. Bifurcation diagram of the instantaneous
turbulent fraction $F_t$ as a function of $\varepsilon$.} \label{F15}
\end{figure}

\begin{figure}[ht]
 \noindent
 \begin{minipage}[b]{.36\linewidth}
 \begin{center}
 \epsfig{file=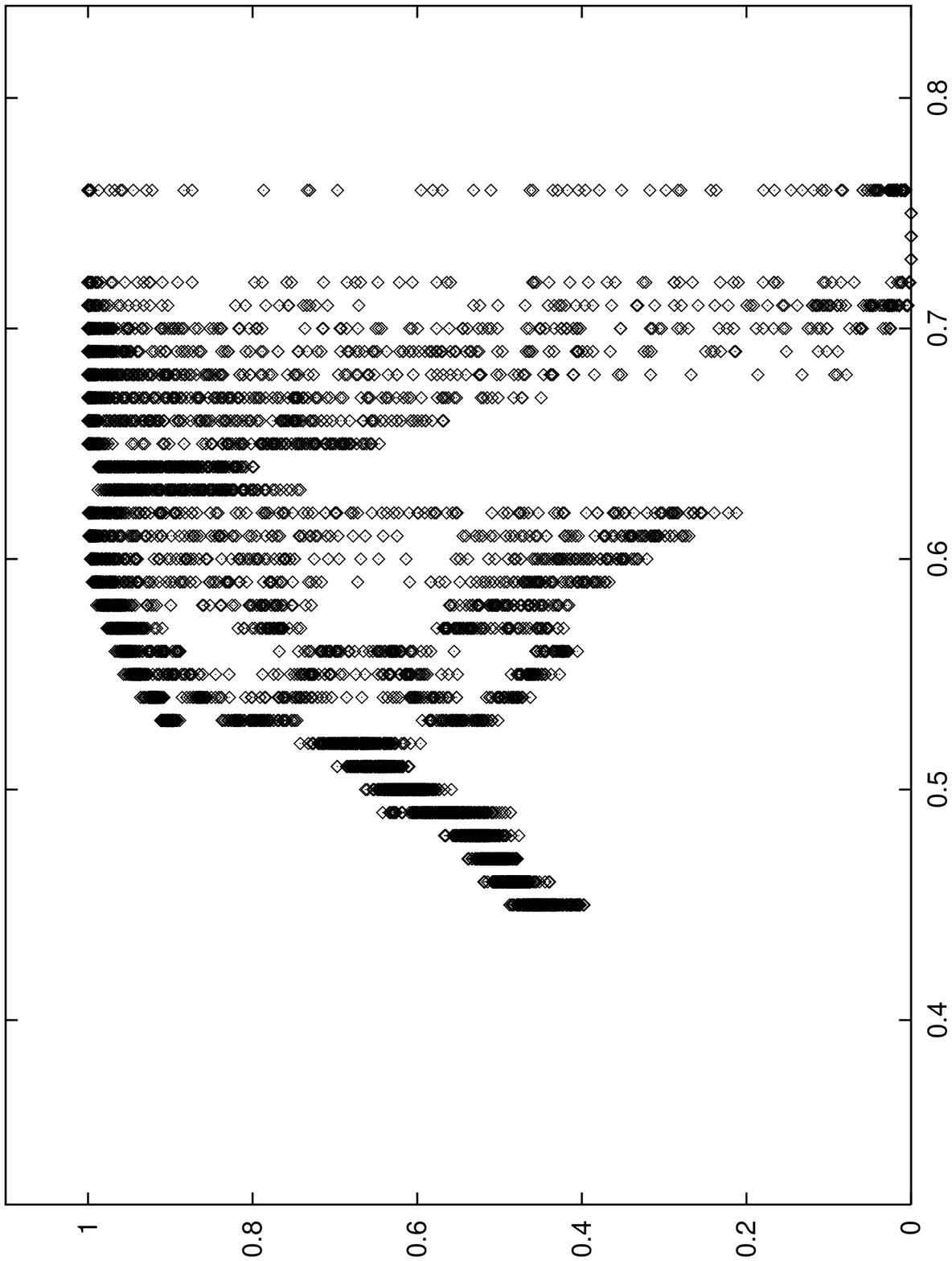, angle= 0}
 \end{center}
\end{minipage}
\caption{  Model A. $\mathbb{G}(10^4,30)$. Bifurcation diagram of $F_t$ vs.
$\varepsilon$.} \label{F16}
\end{figure}

In Figs.~(\ref{F15}) and (\ref{F16}) we show the bifurcation diagram of the
instantaneous turbulent fraction $F_t$ as a function of the coupling $\varepsilon$ for
RCMN induced by the random graphs $\mathbb{G}(10^4,25)$ and $\mathbb{G}(10^4,30)$,
respectively. Figures~(\ref{F15}) and (\ref{F16}) reveal a bifurcating band structure
for the range of coupling corresponding to the observed large fluctuations in $\langle
F \rangle$, reminiscent of the pitchfork bifurcations of unimodal maps.

\begin{figure}[ht]
 \noindent
 \begin{minipage}[b]{.36\linewidth}
 \begin{center}
 \epsfig{file=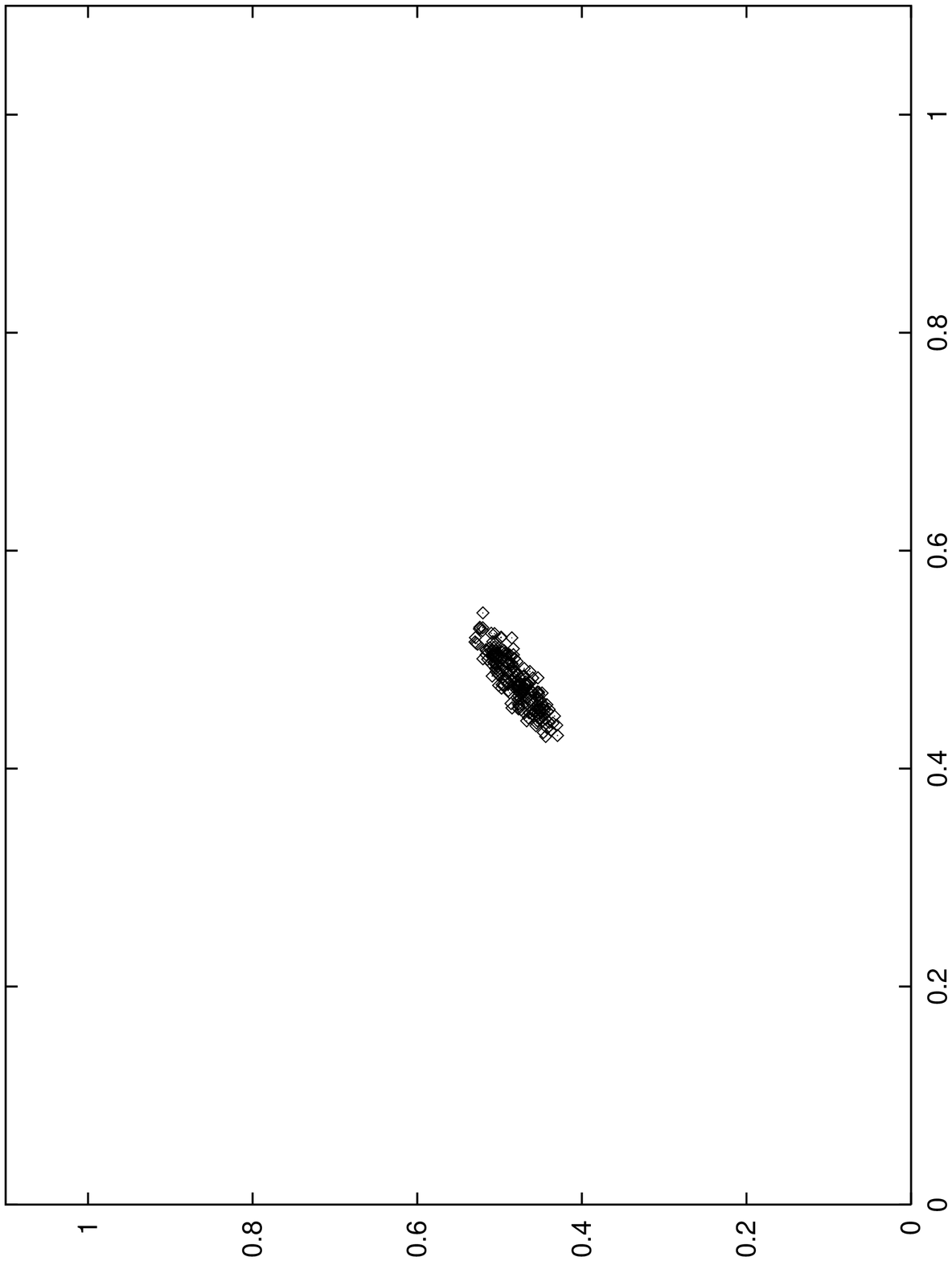, angle= 0}
 \end{center}
\end{minipage}
\caption{Return map $F_{t+1}$ vs. $F_t$ with fixed $r=3$ and $\varepsilon=0.44$, for
Model A, $\mathbb{G}(10^4,25)$. The figure shows the uniqueness of the Gibbs state for
$\varepsilon<1/2$.} \label{F12}
\end{figure}

\begin{figure}[ht]
 \noindent
 \begin{minipage}[b]{.36\linewidth}
 \begin{center}
 \epsfig{file=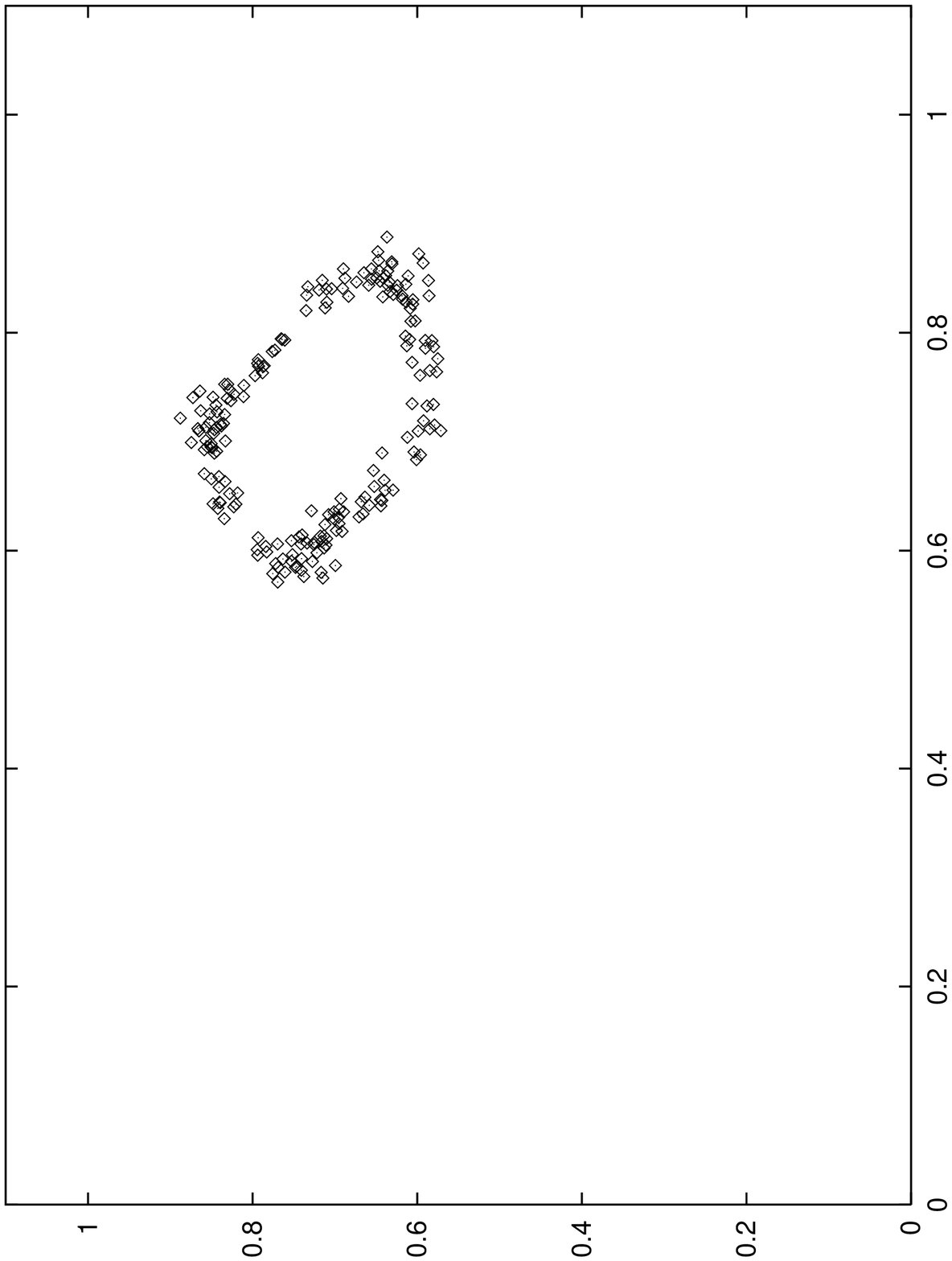, angle= 0}
 \end{center}
\end{minipage}
\caption{Return map $F_{t+1}$ vs $F_t$. Model A, $\mathbb{G}(10^4,25)$; parameters are
$r=3$ and $\varepsilon=0.54$. A $3$-periodic collective motion.} \label{F13}
\end{figure}

\begin{figure}[ht]
 \noindent
 \begin{minipage}[b]{.36\linewidth}
 \begin{center}
 \epsfig{file=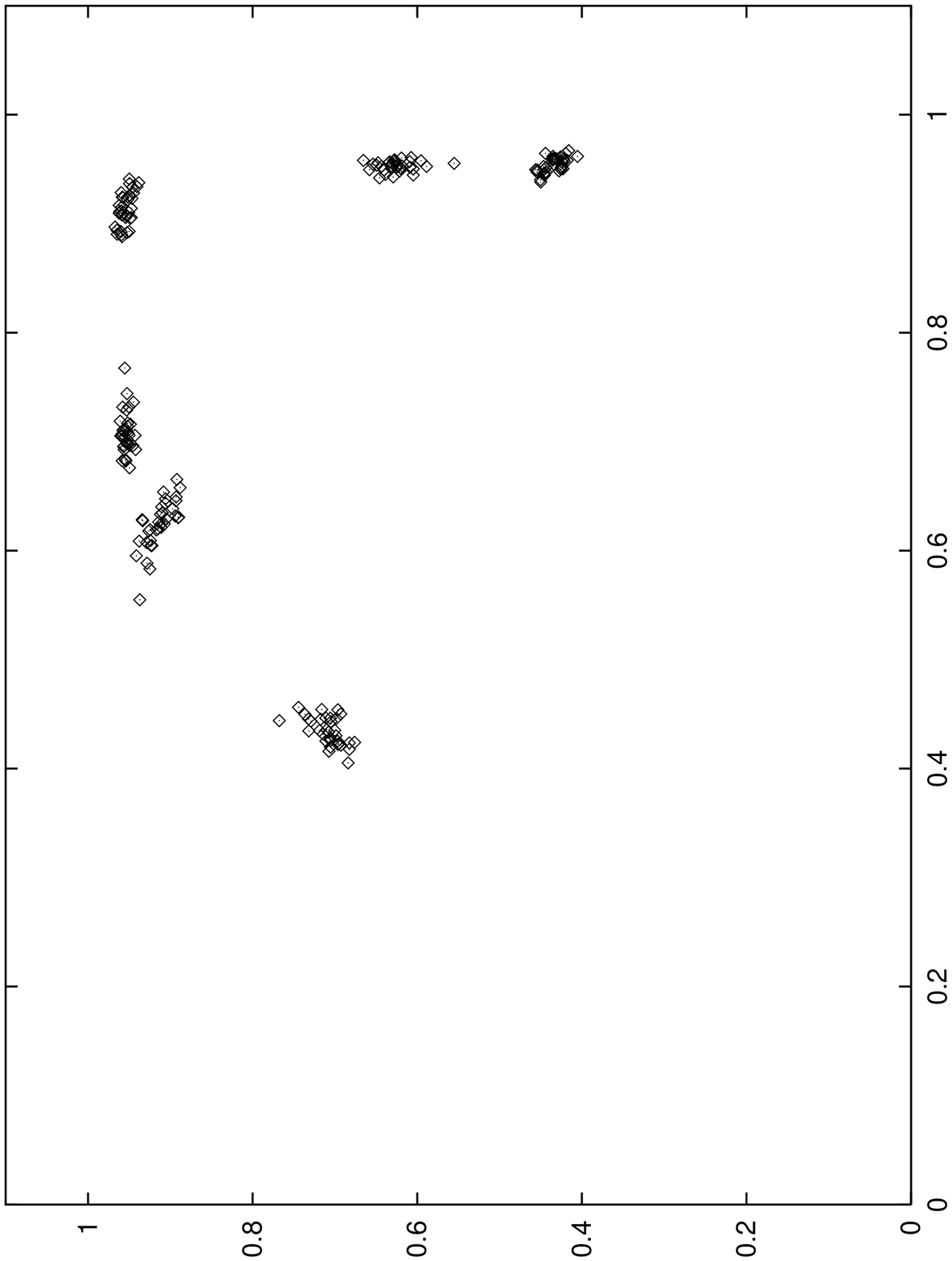, angle= 0}
 \end{center}
\end{minipage}
\caption{ Model A. $\mathbb{G}(10^4,30)$. A $6$-periodic
 collective behavior; $r=3.0$ and $\varepsilon=0.56$.}
\label{F14}
\end{figure}

The return maps at different values of the coupling $\varepsilon$ manifest the
collective nontrivial behavior in the network. The return maps $F_{t+1}$ vs. $F_t$ for
the network $\mathbb{G}(10^4,25)$ show that before the onset of bifurcations, the
sustained turbulent state in the system corresponds to a fixed point with normal
statistical fluctuations, as seen in Fig.~(\ref{F12}). For $\varepsilon=0.54$ in
Fig.~(\ref{F13}), the turbulent fraction shows a period three motion. Other nontrivial
collective states can be observed at different parameter values and for random
networks with different values of $k$. For example, Fig.~(\ref{F14}) shows that the
instantaneous turbulent fraction $F_t$ displays a period-six collective behavior in a
RCMN spanned by the random graph $\mathbb{G}(10^4,30)$ at $\varepsilon=0.56$ and
$r=3.0$.


For large enough connectivities $k$, a relaminarization process is observed in the
systems. That is, at some $\varepsilon'_c>\varepsilon_c$ the mean turbulent fraction
again vanishes, establishing a well defined window of spatiotemporal intermittency.
This phenomenon has also been observed in globally coupled Chat\'e-Manneville maps
\cite{CP}. This suggests that the collective properties of randomly coupled map
networks and globally coupled maps are similar.

\begin{figure}[ht]
 \noindent
 \begin{minipage}[b]{.36\linewidth}
 \begin{center}
 \epsfig{file=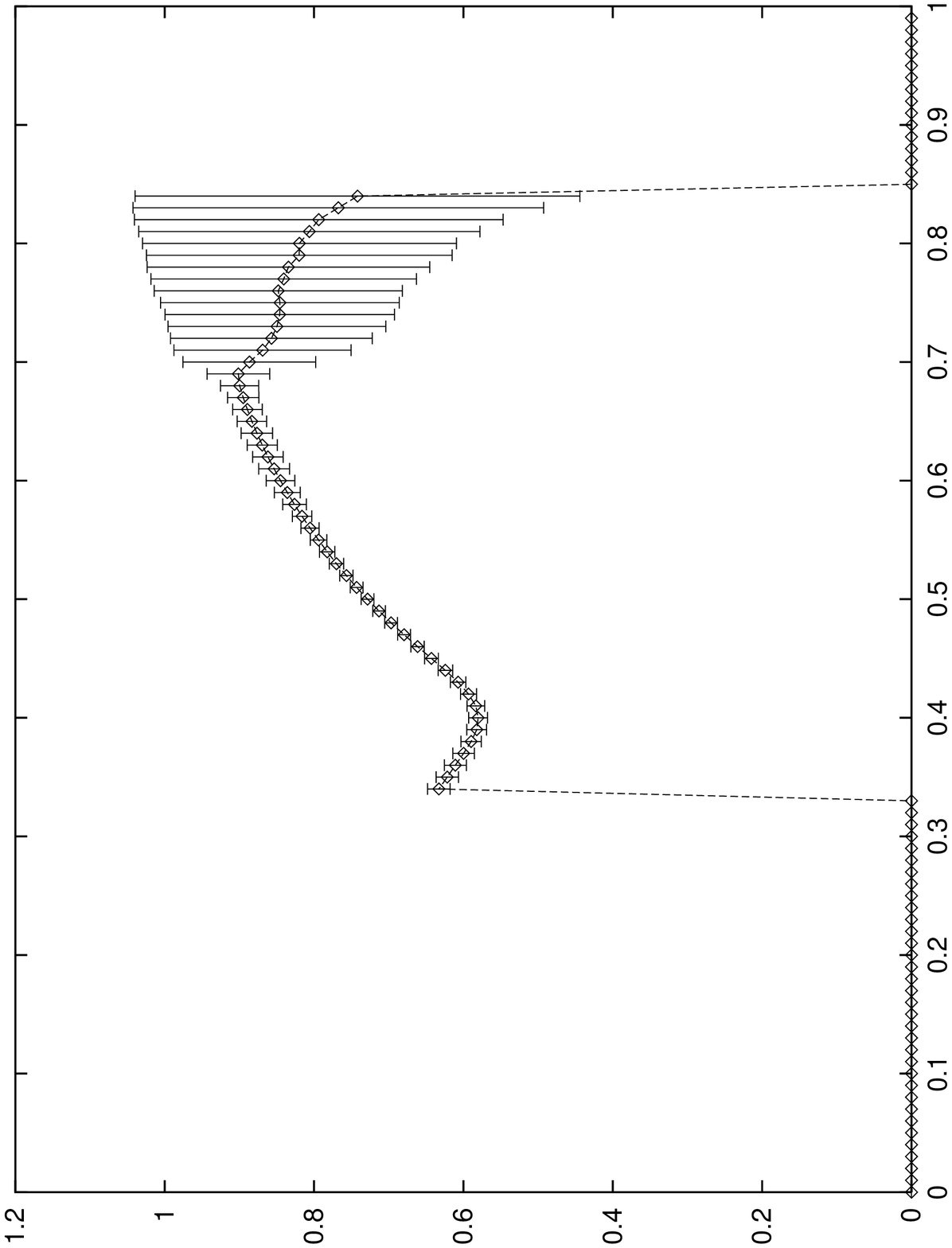, angle= 0}
 \end{center}
\end{minipage}
\caption{  Model A. $\mathbb{G}(10^4,10)$. The average turbulent fraction
$<F>$ vs $\varepsilon$.}
\label{F17}
\end{figure}

\begin{figure}[ht]
 \noindent
 \begin{minipage}[b]{.36\linewidth}
 \begin{center}
 \epsfig{file=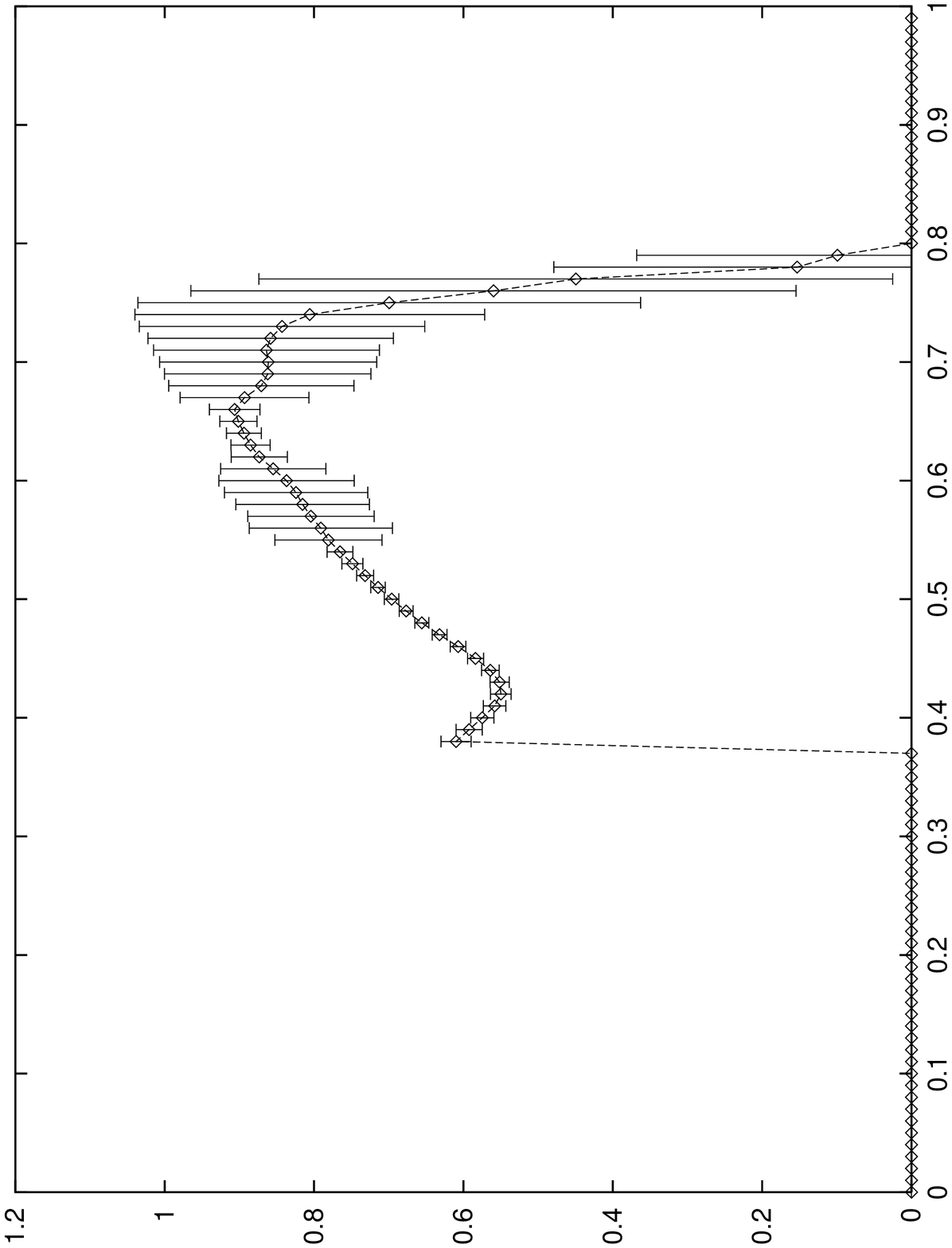, angle= 0}
 \end{center}
\end{minipage}
\caption{  Model A. $\mathbb{G}(10^4,15)$. The average turbulent fraction
$<F>$ vs $\varepsilon$.}
\label{abc}
\end{figure}

\begin{figure}[ht]
 \noindent
 \begin{minipage}[b]{.36\linewidth}
 \begin{center}
 \epsfig{file=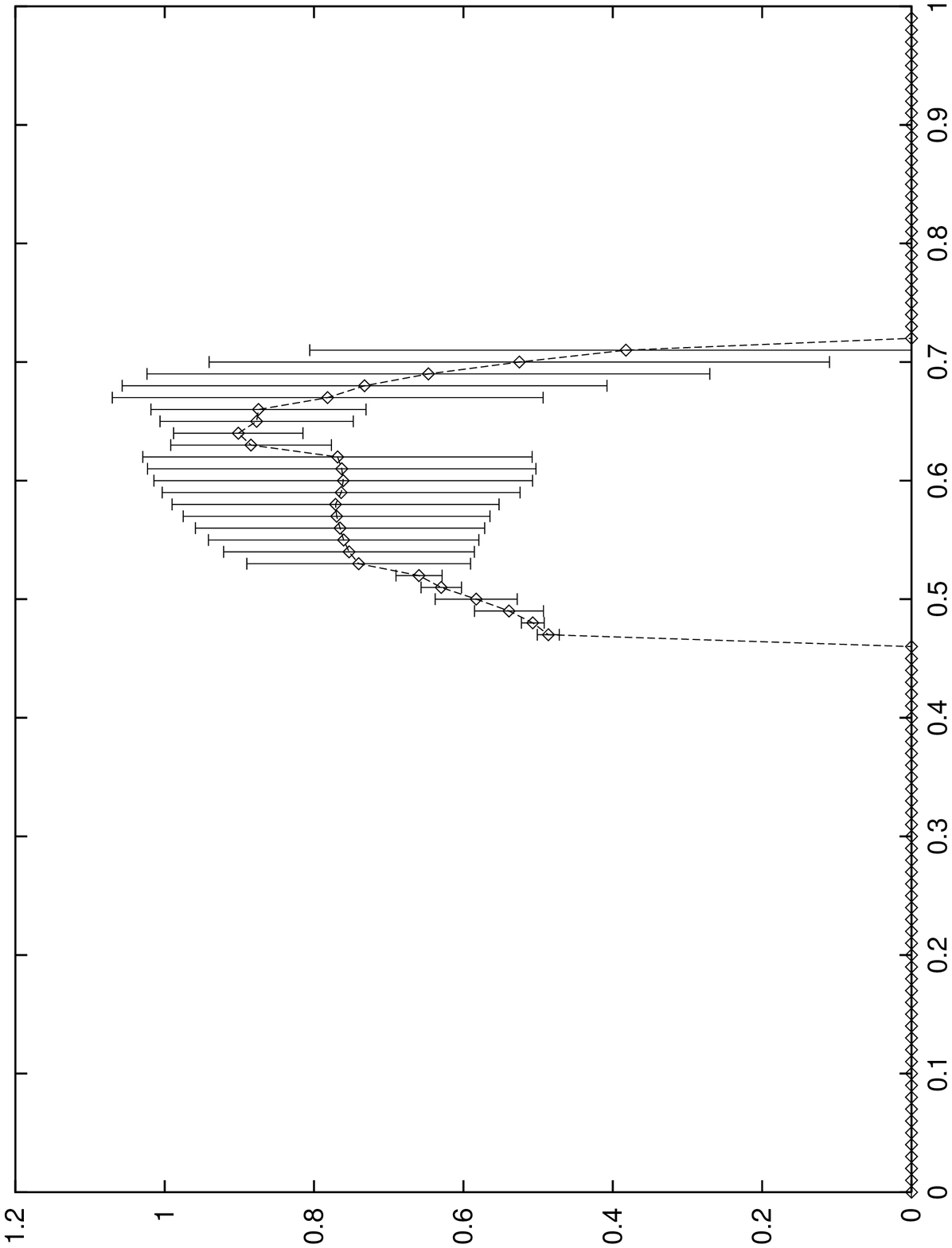, angle= 0}
 \end{center}
\end{minipage}
\caption{  Model A. $\mathbb{G}(10^4,35)$. The average turbulent fraction
$<F>$ vs $\varepsilon$.}
\label{F18}
\end{figure}

Figure~(\ref{F17}) shows $\langle F\rangle$ vs. $\varepsilon$ for the RCMN induced by
the random graph $\mathbb{G}(10^4,10)$. The turbulent window is established within the
interval $\varepsilon\in[0.33, 0.85]$. Both the forward and backward transitions to
the turbulent state appear as discontinuous jumps in the mean turbulent fraction,
similar to the windows of turbulence in globally coupled maps \cite{CP}. However, for
$k>10$, $\langle F\rangle$ decreases gradually, as shown in Figs.~(\ref{abc}) and
(\ref{F18}). For connectivities $15\leq k\leq 40$, the mean turbulent fraction scales
as $\langle F\rangle\propto (\varepsilon'_c-\varepsilon)^{-\gamma}$ close to the
second critical value $\varepsilon'_c$. The second critical exponent is aproximately
the same for different $k$ and was estimated at $\gamma=0.117\pm 0.003$, for fixed
$r=3$.

\begin{figure}[ht]
 \noindent
 \begin{minipage}[b]{.36\linewidth}
 \begin{center}
 \epsfig{file=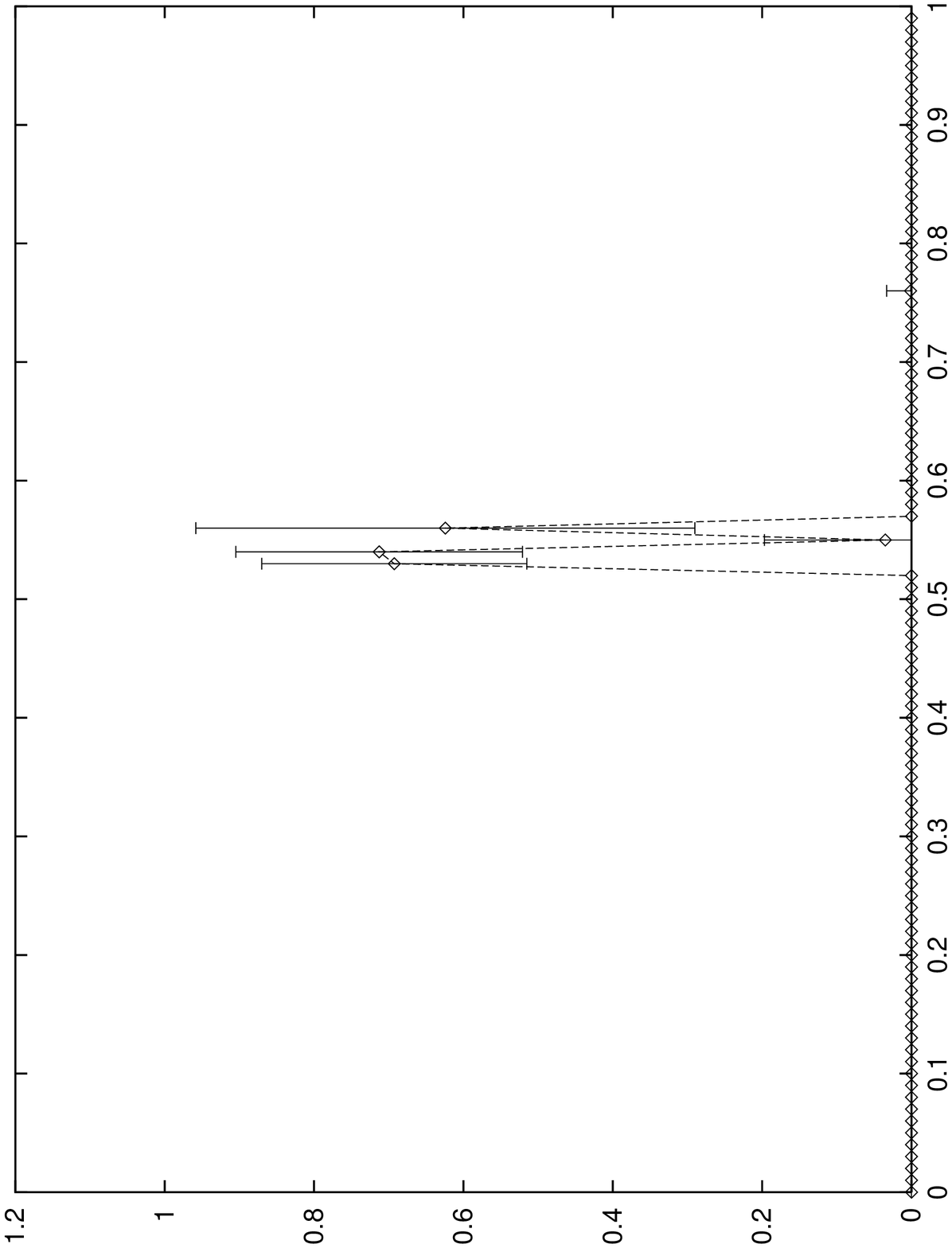, angle= 0}
 \end{center}
\end{minipage}
\caption{The contraction of the turbulent window on the coupling parameter axis as a
function of $k$; $r=3$. a) Model A. b) Model B} \label{2.6}
\end{figure}

As the connectivity $k$ is increased, the windows of turbulence shrink and eventually
disappear, as it can be seen from Figs.~(\ref{F17}), (\ref{abc}), and  (\ref{F18}). We
have plotted the location and the width of the turbulent windows on the coupling
parameter axis as a function of the connectivity $k$ for both model A and model B in
Figs.~(\ref{2.6}a) and (\ref{2.6}b), respectively. In model A, with frozen
connectivity, the turbulent window persists for larger values of $k$.

\section{Probabilistic Geometrical Properties of $\mathbb{G}(N,k)$ }
\label{sec:geom} \noindent

In this section, we take the point of view of random graph theory. The reason for this
is twofold. First, it leads to the understanding of the threshold phenomena occurring
in the transitions to intermittency and relaminarization displayed in the previous
section (see Sec.~\ref{subsec:threshold}). Secondly, the knowledge of local structures
in a random graph and the counting of its small subgraphs allows us to introduce a
notion of \textsl{configuration} which is crucially important  for the thermodynamic
formalism applied to the chaotic coupled maps defined on a random graph (see
Sec.~\ref{sec:TDformRCMN}).

The observations reported in \cite{CM1} and \cite{MM} indicate that the detailed
evolution of dynamical clustering depends crucially on the entire architecture of the
particular network. To define the probabilistic geometrical properties of a random
graph, one has to chose a certain procedure of random graph generation, i.e. a
\textsl{configuration model.} There is a number of constructive procedures
asymptotically almost surely (\textsl{a.a.s.}) leading to $\mathbb{G}(N,k)$. In most
cases, however, these constructions do not give a uniformly distributed random graph,
but it can be checked out if the relevant distributions are contiguous to a uniform
one \cite{RGBook}. In this paper, we follow the configuration model proposed first in
\cite{Ballobas}, and which leads to a uniform distribution of graphs.

Let $N, k\in \mathbb{N}$ be such that $kN$ is even and $k\leq N-1$. The vertex set of
a graph is $\Omega=[N]$. It is natural to define the ``\textsl{in}-" and
``\textsl{out}"-components separately for each vertex as the sets of vertices which
can either be reached or reacheable from a given vertex $\omega\in \Omega$. Let us
arrange that the in-component $\mathcal{I}_t(\omega)\subset \Omega$ is a set of
vertices which are coupled to a given vertex $\omega$ in Eq.~(\ref{cml}) at time $t$.
Consequently, we shall name a set of vertices to which the vertex $\omega$ is coupled
in Eq.~(\ref{cml}) at time $t+1$ as the out-component $\mathcal{O}_t(\omega)\subset
\Omega$.

\subsection{The structure of in-components}
\label{subsec:in}
\noindent

If the connectivity $k$ is fixed, the incoming degree
$s_i=\left|\mathcal{I}_t(\omega_i)\right|$ of the vertex $\omega_i$ in a random graph
is a random variable distributed in accordance with  the Poisson distribution
$\mathrm{Po}(z)=z^ne^{-z}/n!$, where $z=kN/(N-1)$ is the average number of incoming
links \cite{RGBook}, \cite{Newman}. With respect to the backward time direction, the
properties of $\mathbb{G}(N,k)$ are equivalent to those of a uniform directed random
graph $\mathbb{G}(N,kN/2)$.

If $k$ is small and independent of $N$, \textsl{a.a.s} all components of
$\mathbb{G}(N,kN/2)$ are trees or unicyclic, the largest of which have $ O(\log N)$
vertices. As the connectivity
 approaches $k=2$, very quickly all the largest
components merge into one giant component roughly of $O( N^{2/3})$ vertices (see
Fig.~\ref{F4}). The size distribution of remaining small clusters behaves as
$P_\mu\propto \mu^{-3/2}\exp(-\mu)$ \cite{Newman}. Then, another jump in the size of
giant component occurs from $O( N^{2/3})$ to roughly $ O( N)$. This phenomenon of a
``double jump" in the evolution of $\mathbb{G}(N,cN)$ was firstly discussed in
\cite{ER60}.

Note that the appearance of the giant component at $k=2$ however, does not guarantee
that there are no isolated vertices in the graph, and that each vertex can be
reachable from a given one. In fact, as $k=2,$ the random graph consists of a number
of small disjoint clusters of  sizes $m\ll N$.

\subsection{The configuration model and subgraphs classification}
\label{subsec:classif}
\noindent

Next we study the graph following the forward traversal of edges that corresponds to
the natural (forward) lapse of time. In the constructive procedure, we associate the
disjoint $k$-element sets $\mathcal{O}_t(\omega)$ to each element $\omega\in \Omega$
such that $\mathbb{W}_t= \Omega\times \mathcal{O}_t(\omega)$.  The points in
$\mathbb{W}_t$ are the outgoing tails, $|\mathbb{W}|= (kN-1)!!=(kN)!/2^{kN/2}(kN/2)!$.
Then, a configuration $\Theta_t$ is a partition of $\mathbb{W}_t$ into $kN/2$ directed
pairs which we call the outgoing edges. The natural projection $\Pi_t:\mathbb{W}_t \to
\Omega $ projects each configuration $\Theta_t$ to a directed multigraph
$\pi(\Theta_t)$. If $\pi(\Theta_t)$ lacks loops and multiple edges, it is equivalent
$\mathbb{G}(N,k)$.

One should note that if the latter condition on $\pi(\Theta_t)$ being a simple graph
is omitted, we arrive at the model $\mathbb{G}^*(N,k)$ discussed in \cite{Gade}. The
crucial point concerning to $\mathbb{G}^*(N,k)$ is that it does not have a uniform
distribution over all multigraphs on $\Omega$ since different multigraphs arise from
different numbers of configurations (yielding the additional factors of $1/2$ for each
loop and $1/m!$ for each edge of multiplicity $m$). Nevertheless, as $N\to \infty$ any
property that holds \textsl{a.a.s} for $\mathbb{G}^*(N,k)$  also holds \textsl{a.a.s}
for $\mathbb{G}(N,k)$.

With respect to the forward traversal of edges, the random graphs appearing in the
above procedure are the $k$-regular directed random graphs. A cursory observation of
Fig.~\ref{F4} convinces one that such a graph comprises of a set of typical subgraphs.
A standard ground for their classification is given by an \textsl{excess}
\cite{RGBook}. A component $\mathcal{H}$ of a graph is an $\ell$-\textsl{component} if
it has $K>0$ vertices and $K+\ell$ edges, where $\ell$ is the excess of $\mathcal{H}$.
Note that for any connected component $\ell\geq -1$. $\ell=-1$ only for tree like
components that is a finite sequence of edges $(\omega_i,\omega_{i+1})$ such that
$\mathcal{O}_t(\omega_i)=\mathcal{I}_t(\omega_{i+1})$ for $1\leq i\leq m-1$, where $m$
is a \textsl{length} of the path.

Each $0-$component is unicyclic, i.e. a path that starts and
terminates at the same vertex. Other complex $\ell-$components
with $\ell>0$ contains at least two simple sub-cycles.

\subsection{The counting of small subgraphs and configuration}
\label{subsec:counting}
\noindent

Let $k=\left|\mathcal{O}_t(\omega)\right|$ and $N=[\Omega].$ Directly from definitions
it follows that the probability to observe any given set of $m$ disjoint directed
edges on $\mathbb{W}$ in a random configuration reads as \cite{RGBook},
\begin{equation}\label{disjoint}
  p_m=\frac{(kN-2m-1)!!}{(kN-1)!!}.
\end{equation}
Let us note that if $m$ is fixed, this probability shows a power law behavior
\begin{equation}\label{pm1}
  p_m\sim_{N\to \infty}(kN)^{-m},
\end{equation}
  otherwise,
  \begin{equation}\label{pm2}
p_m\sim_{kN-m\to \infty} \left(\frac
eN\right)^m\left(k-\frac{2m}{N}\right)^{kN/2 -m} k^{-kN/2}.
\end{equation}
The latter relation follows from (\ref{disjoint}),  the
expression $(n-1)!!=\sqrt{2}n^{n/2}e^{-n/2}(1+\mathrm{O}(1/n)),$
and the Stirling formula.

Let us count  the number $X^{\ell}_m$ of various small $\ell$-components of the length
$m$ (here, `small' means $m\leq N-1$) appearing in $\mathbb{G}(N,k)$. Note that
$X^{\ell>m-1 }_m\equiv 0,$ and $X^{0}_1=0$ is the number of loops. Therefore,
$X^{0}_2$ is the number of simple directed two-vertex cycles, $X^0_3$ is the number of
directed triangles, etc. As $N\to\infty$ in a random graph,  $X^{\ell}_m$ are the
random variables such that their distributions converge jointly in $\mathbb{R}^\infty$
to the Poisson distributions $\mathrm{Po}(\lambda^{\ell}_k)$, where
$\lambda^{\ell}_k=\mathbb{E}X^{\ell}_m$ are the expectations of $X^{\ell}_m$
\cite{RGBook}.

The number of directed path ($\ell=-1$) of length $m$ can be
calculated readily, $\mathrm{Pa}_m=(N)_mk^m\prod_{i=2}^{m} s_{i}
\simeq N^mk^m\prod_{i=2}^{m} s_{i},$ in which $(N)_m$ is the falling
factorial \cite{ff} and $s_i$ is the incoming degree of the
vertex $\omega_i$ . Remember that $s_i$ is a random variable
having a distribution contiguous to the Poisson one. Then,
$\lambda^{(-1)}_m=\mathbb{E}X^{(-1)}_m=p_m\mathrm{Pa}_m=\prod_{i=2}^m
s_{i}\sim  z^{m-1}=k^{m-1}$ as $N\to \infty.$

Analogously, for the number of simple cycles, one obtains
$\lambda^0_m=m^{-1}\prod_{i=1}^m s_i \sim_{N\to \infty} m^{-1}$
$\times(k)^m,$ in which the factor $1/m$ comes from all
permutations of vertex indices within the cycle. Then, for the
number of $1$-component subgraphs, we arrive at $\lambda^1_m=
(m-2)^{-1}(k-1)^m\prod_{i=1}^m s_i$ $\times\prod_{j=1}^{m-1}
(s_j-1)
\sim_{N\to \infty}k^m(k-1)^m(k-2)^{m-1}/(m-2)$ and so on.

Due to properties of the Poisson distribution, one obtains the
following asymptotic relation for the factorial moments (i.e. the
number of ordered pairs $(X^{\ell}_m)_2$, triplets
$(X^{\ell}_m)_3$, quadruplets $(X^{\ell}_m)_4,$ etc.)
\begin{equation}\label{gen}
  (X^{\ell}_{m_1})_{i_1}  (X^{\ell}_{m_2})_{i_2} \ldots
  (X^{\ell}_{m_n})_{i_n} \longrightarrow_{N\to \infty} (\lambda^{\ell}_{m_1})^{i_1}
  (\lambda^{\ell}_{m_2})^{i_2}\ldots
 (\lambda^{\ell}_{m_n})^{i_n}.
\end{equation}
A set of pairs
$\Theta(\mathbb{G})=\left\{m,X^{\ell}_m\right\}_{\ell=-1}^{m-1}$
is the \textsl{configuration} of a graph $\mathbb{G}$.

\subsection{Hamilton Cycles and Perfect Matching}
\label{subsec:hc}
\noindent

Hamilton cycles $H$ are the  directed cycles of length $N$. The
analysis developed in the previous subsection gives for the
expectation number of cycles $m=N$,
\begin{equation}\label{H}
\mathbb{E}H_k=(N-1)!\frac{(kN-2N-1)!!}{(kN-1)!!}\cdot k^N \prod_{i=1}^N s_i .
\end{equation}
If $k=0$ or $1$, then there is no Hamilton cycles in $\mathbb{G}(N,k)$. If $k=2$
Eq.~(\ref{H}) yields
\begin{equation}\label{H2}
  \mathbb{E}H_2=\frac{(N-1)!}{(2N-1)!!}\sim_{N\to \infty}
  \sqrt{\frac{\pi}{N}} \longrightarrow_{N\to \infty} 0.
\end{equation}
Hence there is also \textsl{a.a.s} no Hamilton cycles in
$\mathbb{G}(N,2)$.

As $k \geq 3$, the number of Hamilton cycles in $\mathbb{G}(N,k)$
exhibits a threshold. Namely, in
\begin{equation}\label{Hnthresh}
  \mathbb{E}H_{k\geq 3}\sim_{N\to
  \infty}\sqrt{\frac{\pi}{2N}}\left[\frac{(k-2)^{k/2-1}}{k^{k/2-2}}\right]^N
\end{equation}
the quantity within the square brackets  is greater than $1$ for
any $k\geq 3$, therefore, $ \mathbb{E} H_{k\geq 3}\to
\infty$ as $N\to \infty.$ Therefore, $\mathbb{G}(N,k)$ has lots of
Hamiltonian cycles when $k\geq 3.$ As a matter of fact, it means
that $\mathbb{G}(N,k)$ has no isolated vertices as $k\geq 3$,
i.e. there is a perfect matching which covers every vertex of the
graph.

The case of $k=4$ is of a particular interest since the number of edges $e=2N$, i.e.
the excess is $\ell=N.$ Here we refer to a result of \cite{RGBook} (see also
references therein) about the contiguity of probability distributions defined on a
simple sum of two Hamilton cycles $\mathbb{H}(N)$ and the random graph
$\mathbb{G}(N,4)$,
\begin{equation}\label{2H}
\mathbb{H}(N)+\mathbb{H}(N)\asymp \mathbb{G}(N,4).
\end{equation}
The latter statement means that, as $N\to \infty$, the probability measures defined on
$\mathbb{G}(N,4)$ and on two independent Hamilton cycles $\mathbb{H}(N)+\mathbb{H}(N)$
are mutually absolutely continuous.

\subsection{Sharp and coarse thresholds in RCMN}
\label{subsec:threshold}
\noindent

In the Sec.~\ref{sec:results}, we have encountered a number of
threshold phenomena related to transitions to intermittency and
backward to a laminar state. At the onset of intermittency, i.e.
as $\varepsilon\to\varepsilon_c-$ and $r$ fixed, there is a
monotone increasing property of $\mathbb{G}(N,k)$ to have an
induced turbulent subgraph $G$ which calls for a close attention.
Similarly, as $\varepsilon\to\varepsilon'_c-$, for $r$ fixed, one
can define a monotone decreasing property of having a laminar
subgraph.

We define the intermittency threshold for the RCMN $\mathbb{G}(N,k)$ as follows. Let
us suppose that there are $F_tN$ excited cells in $\Omega$ at time $t$. Consider a
subgraph $G\subseteq\mathbb{G}(N,k)$ such that the vertex set $[F_tN]$ of $G$ is
$[F_tN]\subseteq[N]$ and the edge set
$E\left[G\right]=E\left[\mathbb{G}(N,k)\right]\cap (F_tN)^2.$ We shall call $G$ as the
\textsl{induced turbulent subgraph} of the random graph $\mathbb{G}(N,k).$

Directly from Eqs. (\ref{cml}) and (\ref{map}) one obtains that a site $\omega$ that
is laminar at time $t$ becomes turbulent at time $(t+1)$ if
$x_\omega(t)\in[1,x_m(\omega;t)]$ where $x_m(\omega;t)$ is the maximum value that a
laminar cell may have in order to become turbulent in the next iteration,
$$x_m(\omega;t)=\frac{1-\varepsilon \varphi(\omega;t)}{1-\varepsilon}, $$
 where
$\varphi(\omega;t)=s_\omega^{-1}\sum_{\omega'\in \mathcal{I}_t(\omega)
}f\left(x_{\omega'}(t)\right)$. Therefore, $P\{1\leq x_{\omega}(t)< x_m(\omega;t)\}$
is the probability that $\omega$ becomes turbulent at the next time step.
Consequently, $P\{x_m(\omega;t)/r<x_{\omega}(t)<1- x_m(\omega;t)/r\}$ is the
probability that the cell $\omega$ being turbulent at time $t$ becomes laminar at time
$t+1$.

Let us denote a sequence of probabilities that the site $\omega$
is turbulent at time $t+1$ at different values of coupling
$\varepsilon$ as
$$
\mathfrak{p}(\varepsilon)=P\{1\leq
x_{\omega}(t)<
x_m(\omega;t)\}
$$
$$
\times\left(1-P\{x_m(\omega;t)/r<x_{\omega}(t)<1-
x_m(\omega;t)/r\}\right).
$$
Then we define a limit
$\mathfrak{p}_c=\lim_{\varepsilon\to\varepsilon_c}\mathfrak{p}(\varepsilon).$
For an increasing property of having the induced turbulent
subgraph $G\subseteq \mathbb{G}(N,k),$ a sequence
$\mathfrak{p}(\varepsilon)$ is called a \textsl{threshold} if
\begin{equation}\label{thr1}
  \mathbb{P}\left\{G\subseteq\mathbb{G}(N,k)\right\}=
  \left\{
\begin{array}{cl}
  0, & \mathfrak{p}(\varepsilon)\ll \mathfrak{p}_c  \\
  1, & \mathfrak{p}(\varepsilon)\gg \mathfrak{p}_c.
\end{array}
  \right.
\end{equation}
Furthermore, $\mathfrak{p}_c$ is called a \textsl{sharp} threshold
if for every $\eta>0$,
$\mathbb{P}\left\{G\subseteq\mathbb{G}(N,k)\right\}=0$ as
$\mathfrak{p}\leq (1-\eta)\mathfrak{p}_c,$ and
$\mathbb{P}\left\{G\subseteq\mathbb{G}(N,k)\right\}=1$ as
$\mathfrak{p}\geq (1-\eta)\mathfrak{p}_c,$ otherwise we shall call
$\mathfrak{p}_c$ as a
\textsl{coarse} threshold.

A recent result \cite{Friedgut} establishes that graph properties that depend on the
inclusion of a large subgraph have always sharp thresholds. A monotone graph property
with a coarse threshold may be approximated by the property of containing at least one
of a certain finite family of small graphs as a subgraph. This statement gives us a
key to understanding the nature of the transitions to intermittency occurring in RCMN.

Indeed, as $k=2$, the random graph $\mathbb{G}(N,k)$ consists of merely small
subgraphs. Some of them become turbulent as $\varepsilon\geq \varepsilon_c$,
establishing a coarse threshold. If $k\geq 3$, the random graph $\mathbb{G}(N,k)$ is
connected; moreover, it comprises of a number of Hamilton cycles, and consequently,
the intermittency threshold is sharp. Otherwise, if the connectivity is around $k=10$,
the relaminarization process which starts as $\varepsilon\to \varepsilon'_c-$ appears
as a sharp threshold since, probably, a whole Hamilton cycle becomes laminar at once.
However, for $k$ substantially greater than $10$, the relaminarization process comes
step by step over small subgraphs establishing a coarse threshold.

We conclude this section with a note on a power law for a monotone graph property
close to a threshold value. For a coarse threshold, there are $n\in \mathbb{N}$ and
$\alpha\in \mathbb{R}$ such that $\mathfrak{p}(\varepsilon,t)\asymp n^{-\alpha}.$ More
precisely, there is a partition of $[N]$ into a finite number of sequences $[N_1],
[N_2],\ldots [N_m]$ (i.e., induced subgraphs) and rational numbers $\alpha_1,
\alpha_2, \ldots \alpha_m>0$ such that $\mathfrak{p}(\varepsilon,t)\asymp
n_j^{-\alpha_j}$ for $n_j\in [N_j],$ \cite{RGBook}.

\section{Thermodynamic Formalism  for Coupled Maps on  Random
Networks}
\label{sec:TDformRCMN}
\noindent

In this section, we consider the thermodynamic formalism (TD) approach to the behavior
of coupled maps defined on random networks. TD relies upon a  symbolic representation
for the coupled maps dynamics. The general idea of the approach is to study this
representation via Gibbs states for the $(d+1)$-dimensional system which goes back to
\cite{S} and \cite{R}.

\subsection{The formal definition of randomly coupled map networks.}
\label{subsec:defRCMN}
\noindent

We give a rigorous definition for ensembles of coupled maps defined on a random graph.
Consider a finite set $\Xi \subset \mathbb{Z}$ such that $|\Xi|=\mathcal{N}< \infty$
and $k\in \mathbb{Z}_+$, ${\ }k\leq \mathcal{N}-1$, such that $k\mathcal{N}$ is even.
Following the standard configuration model (Sec.~\ref{subsec:classif}), one associates
the disjoint $k-$element sets $\mathcal{O}_t(\upsilon)$ to each element $\upsilon\in
\Xi.$ As a result, one arrives at the set of outgoing tails $\mathbb{W}_t=\Xi\times
\mathcal{O}_t(\upsilon).$ A partition $\Theta_t$ of $\mathbb{W}_t$ into
$k\mathcal{N}/2$ directed pairs which we call the outgoing edges. Then the natural
projection $\pi\left(\Theta_t(\mathbb{G})\right)$ is a simple random graph
$\mathbb{G}(\mathcal{N},k)$.

At each node $\varpi\in \Xi,$  we define a local phase space
$\mathrm{X}_{\varpi}$ with an uncountable number of elements. The
global phase space
$\mathrm{M}_{\mathbb{G}(\mathcal{N},k)}=\Pi_{\varpi\in
\Xi}\mathrm{X}_{\varpi}$ is a direct product of local phase spaces such
that a point $\mathrm{x}\in
\mathrm{M}_{\mathbb{G}(\mathcal{N},k)}$ can be represented as
$\mathrm{x}=(x_{\varpi}),$ ${\ }\varpi\in \Xi.$

Let us suppose that there is a subset $\Omega\subset\Xi$ such that $|\Omega|=N\ll
\mathcal{N}$. Consider a subgraph $\mathbb{G}(N,k)\subset \mathbb{G}(\mathcal{N},k)$
such that the edge set
$E\left[\mathbb{G}(N,k)\right]=E\left[\mathbb{G}(\mathcal{N},k)\right]\cap \Omega ^2.$
Then $\mathbb{G}(N,k)$ is a random graph \textsl{induced} by $\Omega$. For each
$\omega\in \Omega$, we denote the local phase space $\mathrm{X}_{\omega}\subseteq
\mathrm{X}_{\varpi}$, and the global phase space
$\mathrm{M}_{\mathbb{G}(N,k)}=\Pi_{\omega\in \Omega}\mathrm{X}_{\omega}$ such that
$\mathrm{M}_{\mathbb{G}(N,k)}\subseteq \mathrm{M}_{\mathbb{G}(\mathcal{N},k)}.$ In
what follows, we denote $\mathrm{M}_{\mathbb{G}(N,k)}$ simply as
$\mathrm{M}_{\mathbb{G}}$ and take the limit $\mathcal{N}\to \infty$.

The \textsl{randomly coupled map network} (RCMN) defined on the random graph
$\mathbb{G}(N,k)$ is a mapping $\Phi_{\mathbb{G}}:\mathrm{M}_\mathbb{G}\to
\mathrm{M}_\mathbb{G}$ which  preserves the product structure,
$\Phi_\mathbb{G}\mathrm{x}=(\Phi_{\omega}x)_{\omega\in \Omega}$, such that
$\Phi_{\omega}:\mathrm{M}_\mathbb{G}\to \mathrm{X}_{\omega}$.

As usual, the mapping $\Phi_\mathbb{G}=C\circ F,$ can be considered as a composition
of the local mapping $(Fx)_\omega=f_\omega(x_\omega)$, which is independent from the
graph topology, $f_\omega:\mathrm{X}_\omega\to \mathrm{X}_\omega$, and the interaction
$(C_\mathbb{G}x)_\omega=g^{\mathbb{G}}_\omega(x)$.

In the framework of thermodynamic formalism, we seek for a symbolic representation for
the dynamic of the coupled map system $(\Phi _\mathbb{G}x )_\omega,$ ${\
}\omega\in\Omega $ on the cylinder $\mathbb{L}=\Omega\times\mathbb{Z}_+$ where
$\Omega\subset\Xi$. Regarding the above definition, models A and B introduced before
can be considered. In Model A, there is the ``frozen disorder" proposed first in
\cite{CM1}, where the configuration $\Theta(\mathbb{G})$ is kept fixed while the map
$\Phi$ is iterating. In Model B, the configuration $\Theta_t(\mathbb{G})$ is changed
at each time step as the map is updated.

Let us note that from the definition of RCMN given above, in the limit $\mathcal{N}\to
\infty$ both  Model A and Model B are not very different. The numbers of small
subgraphs $X^{\ell}_m(\mathcal{N})$ in the entire  random graph
$\mathbb{G}(\mathcal{N},k)$ are random variables fluctuating about their expectation
values $\lambda^{\ell}_m.$ Let us define a discrete time \textsl{random graph process}
$\left\{ \mathbb{G}(\mathcal{N},k) \right\}_{\mathcal{N}}$ which describes the growth
of the entire random graph $\mathbb{G}(\mathcal{N},k)$ as $\mathcal{N}\to \infty.$
This is obviously a Markov process with time running through the discrete set
$\{0,1,\ldots k\mathcal{N}/2\}.$ Hence, one has $t=O(\mathcal{N})$ as $\mathcal{N}\to
\infty.$

The graph $\mathbb{G}(N,k)$ is a small subgraph of $\mathbb{G}(\mathcal{N},k)$ induced
by $\Omega\subset\Xi$. The configuration of $\mathbb{G}(N,k)$ also varies as
$\mathcal{N}$ grows. The numbers of small subgraphs $X^{\ell}_m(\mathcal{N},N)\simeq
X^{\ell}_m(t,N)$ are the Poisson distributed random variables (since
$\mathbb{E}(X^{\ell}_m)_k=(\lambda^{\ell}_m)^k$) such that
$X^{\ell}_m(t,N)\longrightarrow\lambda^{\ell}_m$ as $t\to\infty$ and $N\to \infty.$
Therefore, even in the model of ``frozen disorder", Model A, the actual configurations
would change at each time step.

\subsection{Symbolic  dynamics and Gibbs states for
 the RCMN}
\label{subsec:symbRCMN}
\noindent

Given the configuration of the entire random graph $\Theta_\mathcal{N}=$ $
\pi^{-1}\left(\mathbb{G}(\mathcal{N},k)\right)$, a symbolic dynamics is defined as a
direct product $\mathcal{T}=\pi^{-1}\otimes T$, where $T$ is a semi-conjugacy (since,
in principle, there would be no inverse map on the partition boundaries) to the map
$\Phi$ on $M_\mathbb{G}$ from a subshift $\sigma$ on a symbolic configuration space
$M^{s}_{\Phi}$:
\begin{equation}\label{sd}
  \forall {\xi} \in M^{s}_{\Phi},
  \quad T(\sigma {\xi})= \Phi T ({\xi}),
\end{equation}
and $\pi^{-1}$ is conjugated to a subshift $\tau$ on a random
graph configuration space $\mathbb{W},$
\begin{equation}\label{sd1}
\pi^{-1}(\tau \Theta_\mathcal{N})=
\pi^{-1}(\Theta_{\mathcal{N}+1}).
\end{equation}
Relations (\ref{sd})-(\ref{sd1}) purport a Markov partition $\mathcal{V}_{\xi,\Theta}$
to be defined with an index set $M^s_\mathbb{G}=A^s$ for a finite alphabet $A$. The
simplest possible alphabet would comprise just of two letters, $A=\{0,1\},$ indicating
either ``excited" or ``inhibited" state.

As a result, to any spatio-temporal configuration $x\in
M_\mathbb{G}: x=(x_\varpi),\varpi\in\Xi$, a symbolic code
$\xi=(\xi_t), t\in \mathbb{Z}_+$ is assigned, and
$M^s_{\Phi,\mathbb{G}}$ is the set of all such codes.

The thermodynamic formalism comes about by asking for a
conditional probability distribution on symbolic configurations
defined on a cylinder $\mathbb{L}=\Omega\times \mathbb{Z}_+$, ${\
}\Omega \subset \Xi,$ given a symbolic configuration on the
complement $\mathbb{L}^c=\Omega^c\times \mathbb{Z}_+$, ${\ }
\Omega^c = \mathbb{Z}\setminus \Omega.$ On the uniformly hyperbolic
subsets $K\subseteq M_\mathbb{G}$ these conditional probabilities
are given by the Gibbs states,
\begin{equation}\label{gs}
P\left(\left.\xi_{\mathbb{L}}\right| \xi_{ {\mathbb{L}}^c}\right)
= {Z}^{-1}\exp\left[ -{\mathcal{F}}_{\mathbb{L}}(\xi)\right],
\end{equation}
where $\mathcal{F}_{\mathbb{L}}(\xi),$ ${\ } (\xi_t)_\omega\in
M^s_{\Phi,\mathbb{G}},$ ${\ } (\omega,t)\in \mathbb{L}$ is a part
of the potential
\begin{equation}\label{pot}
{\mathcal{F}} (\xi,\Theta)=\sum_{t\in \mathbb{Z}_+}\log \left| \det [D^{(u)}\Phi]
\left(\mathcal{T}(\sigma^t\xi,\tau^t\Theta)_{(\varpi,t)}\right)\right|,
\end{equation}
which plays the role of a Hamiltonian in statistical mechanics. The Jacobian matrix
$[D^{(u)}\Phi]$ is restricted to a unstable subspace which is the whole tangent space
$TM_\mathbb{G}$ for expanding maps, $N\to \infty.$ We shall drop the index $(u)$ in
the sequel.

The normalization factor $Z$ in (\ref{gs}) is a partition
function,
\begin{equation}\label{pf}
Z = \sum_{\eta\in M_{\Phi,\mathbb{G}}^\mathrm{s}}\exp \left[ -\beta
{\mathcal{F}}(\eta)\right],
\end{equation}
where the sum in (\ref{pf}) is performed over all configurations
$\eta\in M_{\Phi,\mathbb{G}}^\mathrm{s}$ which coincide with
$\xi$ on $\mathbb{L}.$

Although, in general, the feasibility of introducing the thermodynamic formalism
defined in Eqs.(\ref{gs})-(\ref{pf}) for a coupled map lattice in any dimension and
for any values of the coupling strength $\varepsilon>0$ could be questionable,
nevertheless, for a 1D piece-wise linear map this approach is indeed always possible
\cite{Jiang}.

\section{Transitions to Intermittency and Collective Behavior \\  in
the RCMN}
\label{sec:transitions}
\noindent

All information on transitions to intermittency and collective behavior in the RCMN is
contained in the Gibbs potential $\mathcal{F}$ (\ref{pot}). In statistical mechanics
it is somewhat unusual because, even in the uncoupled case, the Gibbs potential has
nontrivial interactions in the time direction \cite{BK}, \cite{JP}. However, a formal
analogy between transitions to spatiotemporal intermittency observed in the RCMN and
phase transitions in uniaxial ferromagnets can be found.

For the coupled map system (\ref{cml})-(\ref{map}), the potential $\mathcal{F}$ is a
function of three external parameters, $\{\varepsilon, r,k\}.$ The hyperbolicity of
phase space means a positivity of all Liapunov exponents in the spectrum,
$\lambda_n>0$. From  direct numerical simulations, it is known that the number of
positive Liapunov exponents for extended chaotic systems scales as the lattice size
\cite{C}, $N_{\lambda_n>0}\sim N.$ This means that in the extended limit $N\to \infty$
the instantaneous turbulent fraction  $F_t= N_{\mathcal{T}}(t)/N$ is a natural order
parameter monitoring the transition to intermittency in a coupled chaotic map system.

Since the matrix element of $[D\Phi]$ in (\ref{pot}) relevant to a
site $\varpi$ being in the turbulent state is proportional to $r$,
the instantaneous turbulent fraction can be simply counted  as
\begin{equation}\label{inst}
  F_t=-\frac{\partial {\mathcal{F}}( r,\varepsilon,k)}{\partial r}.
\end{equation}
The analogy with the uniaxial ferromagnet is the following. Let us assign the
turbulent state to a ``spin up" configuration and the laminar state to a ``spin down".
Then $F_t$ is a spontaneous magnetization in the ferromagnet with the interaction
Hamiltonian (\ref{pot}). The mean turbulent fraction
\begin{equation}\label{<F>}
\langle F\rangle =\frac{\partial \log Z}{\partial r},
\end{equation}
in which $Z$ is the partition function (\ref{pf}), is equivalent to the magnetization
in the ferromagnet.

Carrying on this analogy, one can introduce the Gibbs free energy function
$\mathcal{U}$ for the system of coupled chaotic maps. Let us define a finite piece
$\mathbb{L}_\mathcal{N}\subset\mathbb{L}$ of the cylinder $\mathbb{L}=\Omega\times
\mathbb{Z}_+$ having a volume $\mathbb{V}_\mathcal{N}=N\cdot c\mathcal{N}$ where
$c<\infty.$ One can introduce a restriction $\mathcal{F}_{\mathbb{L}_\mathcal{N}}$ of
the potential $\mathcal{F}_\mathbb{L}$ on the finite cylinder
$\mathbb{L}_\mathcal{N}.$ Then the free energy is $\mathcal{U}=\lim_{\mathcal{N}\to
\infty}\mathcal{F}_\mathcal{N}/\mathbb{V}_\mathcal{N}.$

\subsection{The equation for the free energy function}
\label{subsec:CMmap}
\noindent

Following \cite{BK}, \cite{JP}, and \cite{McKay}, we now apply a simple transformation
to (\ref{pot}). We use the standard relation $\log{\ }\det=\mathrm{Tr}{\ }\log$ which
gives us the following expression for the Gibbs potential
\begin{equation}\label{p1}
{\mathcal{F}}_\mathbb{L} (\left.
\xi\right|_\mathbb{L},\Theta)=\sum_{t\in
\mathbb{Z}_+}
\mathrm{Tr} \left( \log \left| [D\Phi] \right|
\right)_{\omega,\omega} \left(
\mathcal{T}(\sigma^t\xi,\tau^t\Theta)_{(\omega,t)}
\right), \quad (\omega,t)\in \mathbb{L}.
\end{equation}
Since the local map (\ref{map})is 1D, then $\log {D\Phi_\omega}$ can be defined as the
number $\log |D\Phi_\omega|.$ The notation $\left(\log\left|
[D\Phi]\right|\right)_{\omega,\omega}$ indicates the diagonal block corresponding to
site $\omega\in \Omega$. The trace is summed over the whole induced subgraph
$\mathbb{G}(N,k)\subset\mathbb{G}(\mathcal{N},k)$ and is independent of any choices.

Returning to the map Eqs.~(\ref{cml})-(\ref{map}), one can proceed further using the
potential (\ref{p1}). The Jacobian matrix in (\ref{p1}) can be written in the form
$[D\Phi]=\mathbb{U}(\mathbb{I}-\mathbb{U}^{-1}\mathbb{C})$, where $\mathbb{U}$ is a
contribution coming from the \textsl{uncoupled} maps (i.e., the diagonal part of
$[D\Phi]$), and $\mathbb{C}$ comes from the coupling. Then we expand $\mathrm{Tr}
\log\left|[D\Phi]\right|=\mathrm{Tr} \log
|\mathbb{U}|-\sum_{s>0}\mathrm{Tr}\left[(\mathbb{U}^{-1}\mathbb{C})^s/s\right].$

The entries $\log |U_{\omega}|$ can take two different values depending on whether
site $\omega$ is laminar or turbulent: $\log| U_{\omega}|=\log(1-\varepsilon)$ if
$1\leq x_\omega\leq r/2$; otherwise,  $\log| U_{\omega}|=\log(1-\varepsilon)+\log r$,
if
 $0\leq x_\omega < 1$. Suppose that there are $NF_t$
turbulent sites in the  graph $\mathbb{G}(N,k)$ induced by $\Omega$ at time $t$.
Therefore, $\mathrm{Tr}\left( \log
|\mathbb{U}_\omega|\right)_{(\omega,\omega)}=N\log(1-\varepsilon) +NF_t\lambda_0$,
where $\lambda_0=\log r$ is the Liapunov exponent of the uncoupled, single map
Eq.~(\ref{map}).

The series $\sum_{s>0}(\mathbb{U}^{-1}\mathbb{C})^s/s$ deserves careful consideration.
It is easy to check that
\begin{equation}\label{u-1c}
\mathbb{U}^{-1}\mathbb{C}=\frac{\varepsilon}{k(1-\varepsilon)}\mathbb{A}_{\omega\omega'},
\end{equation}
in which $\mathbb{A}_{\omega\omega'}$ is the 'weighted' adjacency
matrix of the  random graph $\mathbb{G}(N,k)$ such that
\begin{equation}\label{adjmatrix}
  \mathbb{A}_{\omega\omega'}=\left\{
\begin{array}{cl}
  0, & \omega {\ }\mathrm{and} {\ } \omega' {\ }\mathrm{are {\ } not {\ } coupled,} \\
  1 & \omega {\ }\mathrm{and} {\ } \omega' {\ }\mathrm{are {\ } in {\ }the{\ }same{\ }state,}  \\
  r &  \omega{\ }\mathrm{is {\ } turbulent, {\ }} \omega'{\ }\mathrm{is {\ }laminar,} \\
  1/r &\omega{\ }\mathrm{is {\ } laminar, {\ }} \omega'{\ }\mathrm{is {\ }turbulent.}
\end{array}
  \right.
\end{equation}
The matrix $\mathbb{A}_{\omega\omega'}$ contains data of both
topological as well as dynamical configurations of the coupled
map system defined on $\mathbb{G}(N,k)$. We denote the adjacency
matrix of the graph $\mathbb{G}(N,k)$ as $\mathrm{A}$ with
entries $A_{ij}=0 {\ }\mathrm{or} {\ } 1$.

The number of cycles in $\mathbb{G}(N,k)$ is of  crucial importance. Let us recall
that $\mathrm{Tr}(\mathrm{A}^s)=X^0_s$, i.e. the total number of cycles of the length
$s=\{1,\ldots N\}$ in a graph with the adjacency matrix $\mathrm{A}$ \cite{Codbook}.
While interested in the number of cycles $X^0_s$ in the random graph
$\mathbb{G}(N,k)$, we note that in the matrix $\mathbb{A}$, for each entry
proportional to $r$ contributing in a cycle, there is always present the entry $1/r$
such that they are divided out. Therefore, one can prove that
$\mathrm{Tr}(\mathbb{A}^s)=\mathrm{Tr}(\mathrm{A}^s)=X^0_s.$ We recall that $X^0_1=0$
is the number of loops which are ruled out in our model.

Collecting  the results of the present subsection and taking the expression
Eq.~(\ref{<F>}) for the mean turbulent fraction $\langle F\rangle$ together with its
formal definition Eq.~(\ref{<<F>>}) into account, we arrive at the equation for the
free energy $\mathcal{U}$ of the randomly coupled CM map system,
\begin{equation}\label{p3}
\mathcal{U}=
\log(1-\varepsilon)+\lambda_0\frac{\partial\log
Z }{\partial r } -\frac 1N\sum_{s>1}^{N}\frac 1s
  \left[\frac{\varepsilon}{k(1-\varepsilon)}\right]^sX^0_s,
\end{equation}
where $Z$ is the partition function Eq.~(\ref{pf}).

The first term in the Eq.~(\ref{p3}) is irrelevant  to either the coupled maps
dynamics   or  the random network topology. The second one is a cumulative
contribution from all  chaotic configurations allowed $\eta\in
M_{\Phi,\mathbb{G}}^\mathrm{s}$ which coincide with a given one on the cylinder
$\mathbb{L}.$ Finally, the third term represents  a contribution from the topology of
the random network.

Equation~(\ref{p3}) can hardly be solved explicitly. Nevertheless, some limiting
solutions of (\ref{p3}) can easily be found.

\subsection{Transitions to the spatio-temporal intermittency and relaminarization}
\label{subsec:intermitt}
\noindent

Transitions to the spatio-temporal intermittency and relaminarization  deserve the
name of phase transitions since the behavior of the mean turbulent fraction $\langle
F\rangle$ close to the critical values of coupling $\varepsilon_c$ resembles the
behavior of thermodynamical quantities close to a critical point.

For the connectivity $k=2,$ the coupled maps defined on  either regular lattice,
\cite{CM1}, or at random exhibit a scaling behavior $\langle F\rangle\propto
(\varepsilon-\varepsilon_c)^\beta$ with some critical exponent $\beta$ that is typical
for a \textsl{second order} phase transition. In contrast, the transition between
laminar states and turbulence for either  the RCMN with $k\geq 4$ or the globally
coupled maps, \cite{CP}, appears as a discontinuous jump in $\langle F\rangle$, a
feature associated to \textsl{first order} phase transitions.

For a backward transition from turbulence to a uniformly laminar states, the situation
is different. For minimal connectivities, in both regular coupled maps with local
interactions \cite{CM1} and randomly coupled maps, such a transition does not occur
for any $\varepsilon< 1$. For either  globally coupled maps \cite{CP}, or randomly
coupled maps with connectivities around $k=10$, this transition appears as a
discontinuous jump. However, for randomly coupled maps with connectivities $k>10$,
this transition follows a power law behavior in $\langle F\rangle$ with another
critical exponent $\gamma$. The data convincingly show that the formal analogy with
phase transitions occurring in statistical mechanics cannot provide us the complete
and adequate classification for ``critical" phenomena in the RCMN.

It is quite obvious  that in a laminar domain $\mathcal{L}$ of the space of external
parameters $\mathbb{D}\equiv\{\varepsilon,k, r\}$, the probability $P(\xi)$ (\ref{gs})
to observe a symbolic chaotic configuration $\xi $ on $\mathcal{L}$ is always $P=0$
for any $\xi.$ Therefore, one could expect that the potential Eq.~(\ref{pot}) over the
laminar domain $\mathcal{L}$ is $\left.\mathcal{F}\right|_\mathcal{L}=-\infty.$

When the coupling is small $\varepsilon\ll 1$, the influence of random graph topology
vanishes. Therefore, the series term in the r.h.s. of  Eq.~(\ref{p3}) can be neglected
in this case. Since $\mathcal{F}=\mathbb{V}\mathcal{U}$ and the cylinder volume
$\mathbb{V}$ is taken to be infinite, one can see that the Gibbs potential becomes
$\mathcal{F}=-\infty$, while
\begin{equation}\label{until}
  \langle F\rangle_{\mathrm{min}}<\frac{\left|\log(1-\varepsilon_c)\right|}{\log r}.
\end{equation}
This expression relates the minimal mean turbulent fraction
$\langle F\rangle_{\mathrm{min}}$ which can arise at the onset of
intermittency for a given value $\varepsilon_c$.

For $\varepsilon$ that is not small, the random topology of the network becomes
significant. Instead of (\ref{until}),  one obtains
\begin{equation}\label{until2}
  \langle F\rangle_{\mathrm{min}}<\frac{\left|\log(1-\varepsilon_c)\right|}{\log r}+
  \frac 1{N\log r}\left|
  \sum^N_{s>1}\frac 1{s^2}\left[\frac{\varepsilon_c}
  {1-\varepsilon_c}\right]^sX_s^0\right|.
\end{equation}
Let us consider the series term in Eq.~(\ref{p3}). In a random graph
$\mathbb{G}(N,k)$, the numbers of cycles $X^0_s$ ar the Poisson distributed random
variables $\mathrm{Po}(\lambda^0_s)$ with means $\lambda^0_s=k^s/s$. If the number of
vertices $N$ in the graph is very large, one can replace the $X^0_s$ in (\ref{p3}) by
their expectations $\lambda^0_s$. Consequently, for $N$ large, one arrives at the
following expression
\begin{equation}\label{hypergeom}
  \frac 1N\sum^N_{s>1}\frac 1{s^2}\left[\frac{\varepsilon}{1-\varepsilon}\right]^s =
\frac{\varepsilon^2}{4N(1-\varepsilon)^2}\cdot
F\left([1,2,2],[3,3],\frac{\varepsilon}{1-\varepsilon}\right)
\end{equation}
$$ -\frac{\varepsilon^{N+1}}{N(N+1)^2(1-\varepsilon)^{N+1}}\cdot
F\left([1,N+1,N+1],[2+N,2+N],\frac{\varepsilon}{1-\varepsilon}\right), $$
where
$F\left([a],[b],x\right)$ is the generalized hypergeometric function, in which $[a]$
and $[b]$ are the sets of parameters.

The behavior of the term (\ref{hypergeom}) on the parameter $\varepsilon$ strongly
depends of the random graph topology at given $k$. For random graphs with minimal
connectivities $k=2$, the random variables counting the number of small cycles whose
length $s$ exceeds some maximal length $s>s_{\mathrm{max}}$ is $X^0_s=0$ . The
quantity $s_{\mathrm{max}}$ cannot exceed the size of the giant component of the
random graph $\mathbb{G}(N,k)$, but it is actually much smaller. Hence, the effective
summation in the r.h.s. of Eqs.~(\ref{p3}) and (\ref{until2}) is up to
$s_{\mathrm{max}}\ll N.$ The contribution to (\ref{p3}) coming from the term
(\ref{hypergeom}) slowly increases with $\varepsilon$  as $\varepsilon<1/2$, enhancing
the critical value $\varepsilon_c$ for the intermittency onset.

The series term (\ref{hypergeom}) plays a crucial role in the transition to a
uniformly laminar state. As $k\gg 1$ and $\varepsilon/(1-\varepsilon)\gg 1,$ the value
of the sum abruptly jumps at some value $\varepsilon>1/2$ to numbers much larger than
$ N$. Therefore, the major contribution to the sum comes from the Hamilton cycles $s=
N$. The window of turbulence closes when $\mathcal{U}<0$.

We conclude this subsection with the notion that no turbulent window would appear in
the system for $k\geq 100$. In previous sections, we have shown that for connectivity
values $k\gg 1$, the relevant random graph $\mathbb{G}(N,k)$ has no isolated vertices,
i.e. for every ordered pair of vertices $\omega_i$ and $\omega_j$ there is a path in
$\mathbb{G}(N,k)$ starting in $\omega_i$ and terminating at $\omega_j$. Following the
traditional terminology, we shall call such a graph as \textsl{irreducible}.
Consequently, the adjacency matrix $\mathrm{A}$ of the irreducible graph satisfies the
following property: for each pair of indices $(i,j)$ there exists some $n\geq 0$ such
that $A^n_{ij}>0$ \cite{Codbook}. The typical length of the shortest path between two
arbitrary vertices in an irreducible random graph is $d_{\omega_i\omega_j}=\log N/\log
k$ \cite{Newman}.

Let us consider the adjacency matrix $\mathrm{A}$ of the graph irreducible
$\mathbb{G}(N,k)$. Define a \textsl{period} $p_\omega$ of a node $\omega\in\Omega$ as
the greatest common devisor of those integers $n\in \mathbb{N}$ for which
$(\mathrm{A}^n)_{\omega\omega}>0.$ Then the period $p_\mathrm{A}$ of the matrix is the
greatest common divisor of the numbers $p_\omega$, ${\ }\omega\in \Omega,$
\cite{Codbook}. It is to be noticed that in the model in question $p_\mathrm{A}=2$,
since loops are ruled out. We shall call the nodes $\omega_i$ and $\omega_j$ as
\textsl{period equivalent}, if  the length of  path between them,
$d_{\omega_i\omega_j}$ is divisible by $p_\mathrm{A}$. One can see that as $k=100$,
for $N=10^4,$ all the nodes of the network are period equivalent and, therefore,
synchronized.

\subsection{Transitions to collective behavior in RCMN}
\label{subsec:cbtransitions}
\noindent

Within windows of turbulence, the Gibbs states given by the formula (\ref{gs}) are not
trivial, and the Gibbs potential $\mathcal{F}$ remains finite. A phase transition to
the collective behavior occurs in the system of coupled maps in the thermodynamic
limit $N\to\infty$ when the Gibbs state (\ref{gs}) is not unique, i.e. there are
several (at least two) different Gibbs states with respect to the potential
$\mathcal{F}$ defined on symbolic configurations in $M^s_\Phi$.

In the context of the thermodynamic formalism applied to CML, this idea has been
proposed in \cite{McKay}. For each topologically mixing component of the uniformly
hyperbolic subset $K\subseteq M_{\Phi,\mathbb{G}}^\mathrm{s}$, there is precisely one
asymptotic probability distribution which is called the SRB measure (after
Sinai-Ruelle-Bowen) on the attractor. For $n$-periodic components of $K$, however,
there are $n$ different Gibbs states relevant to $n$-cycling through the
subcomponents.

In this subsection, we demonstrate that, in the thermodynamic limit $N\to \infty$, the
Gibbs potential and consequently the free energy function are multivalued functions as
$\varepsilon>1/2$.

Let us compute the sum in the r.h.s. of the Eq. (\ref{p3}) as
$N\to \infty$. In this case, one obtains
\begin{equation}\label{polylog}
\lim _{N\to\infty}\sum_{s>1}^{N}\frac 1s
\left[\frac{\varepsilon}{k(1-\varepsilon)}\right]^sX^0_s=
\mathrm{Polylog}\left(2,\frac{\varepsilon}{1-\varepsilon}\right)-
\frac{\varepsilon}{1-\varepsilon},
\end{equation}
where $\mathrm{Polylog}(2,\alpha)$ is the polylogarithm function.

The point $\varepsilon = 1/2$ is a branch point for all branches of the polylogarithm
function. The branch cut can be taken to be the interval $(1,\infty)$.  The point
$\varepsilon = 0$ is also a branch point, and the branch cut is taken to be the
negative real axis. The principal branch is given on the unit disk by the series
$$\mathrm{Polylog}(2,\alpha)=\sum_{k=1}^{\infty}\frac{\alpha^k}{k^2}, $$
 and there is
the general formula for the $(n,m)-$th branch of $\mathrm{Polylog}(2,\alpha)$,
 $$
\mathrm{Polylog}(2, \alpha)+2i\pi n\log(\alpha) -4\pi^2 nm, \quad n,m \in \mathbb{Z},
$$
where $\log(\alpha)$ means the principal branch of the logarithm \cite{Lewin}.

If we now introduce the result (\ref{polylog}) into Eq.~(\ref{p3}) and neglect terms
$O(1/N)$ as $N\to \infty$, we arrive at the expression
\begin{equation}\label{Umulti}
\mathcal{U}_{y,m}= \log(1-\varepsilon)+\lambda_0\frac{\partial\log Z }{\partial r }
-4\pi^2ym -2\pi y i \log\left(\frac{\varepsilon}{1-\varepsilon}\right) +
\mathrm{O}\left(\frac 1N\right),
\end{equation}
which reveals the two-parameters set of branches for the free energy function
$\mathcal{U}$. Here, we have introduced  $m\in \mathbb{Z}$, and $y\equiv  n/N \in
\mathbb{R}$ as $N\to \infty.$ The discrete parameter $m$ enumerates different bands of
possible solutions of the Eq.~(\ref{p3}), and the  parameter $y$ (continuous in the
thermodynamic limit) enumerates the different branches of $\mathcal{U}$ within a band.
The behavior prescribed by  Eq.~(\ref{Umulti}) is clearly seen on  Figs.~(\ref{F15})
and (\ref{F16}).

\subsection{The bifurcation route to collective periodic  behavior in RCMN}
\label{subsec:evidence}
\noindent

The graphs presented on Figs.~(\ref{F15}) and (\ref{F16}) essentially resemble the
``bifurcation diagrams"  well known in deterministic chaotic dynamics of  unimodal
maps over the interval $I=[0,1]$ having a negative Schwarzian derivative.

Consider the irreducible random graph $\mathbb{G}(N,k)$ and its adjacency matrix
$\mathrm{A}.$ The Perron-Frobenius theory is applied completely to such an irreducible
matrix. As a particular consequence of this theory concerning the collective behavior
in  the RCMN, one can prove that there  exist two stable fixed points for the map
$\Psi: F_t\to F_{t+1}$ correspondent to either the uniformly synchronized laminar
state or to the sustained fully turbulent state of network. Close to these fixed
points, a map for the instantaneous turbulent fraction $\Psi$ is a polynomial map  in
$F_t.$ Therefore, in the unite interval $I=[0,1]$ it turns to be a unimodal function
having the negative Schwarzian derivative over the whole interval,
\begin{equation}\label{Schwarz}
\mathcal{S}\Psi\equiv \frac{\Psi'''}
{\Psi'} - \frac 32\left(\frac{\Psi''}{\Psi'}\right)^2<0.
\end{equation}
In this case, $\Psi$ displays an infinite sequence of pitchfork bifurcations when the
attractor relevant to the unique Gibbs state losses its stability. An example of such
a map has been presented in \cite{CP}. These bifurcations are actually observable in
some intervals of the parameter values.

In the diagrams shown on Figs.~\ref{F15} and \ref{F16}, the
bifurcation branches  which  draw  away from the stable point
$F_t=1$ terminate soon, while for  the  branches which tend to the
fixed point where the map $\Psi$ is still polynomial, the
consequent pitchfork bifurcations are still observable up to the
very end of the turbulent window.

\section{Conclusion}
\label{sec:conclusion}
\noindent

In this article we have studied the main features associated with transitions to
spatiotemporal intermittency and relaminarization, as well as transitions to
collective behavior occurring in randomly coupled Chat\'{e}-Manneville minimal maps.
Numerical simulations as well as a theoretical framework for these systems have been
presented. We have reviewed and classified previous studies devoted to randomly
coupled maps networks according to the random graphs spanning the networks. We have
studied the probabilistic geometrical properties of the $k-$out model random graphs
$\mathbb{G}(N,k)$. The thermodynamic formalism based on the symbolic representation
for the dynamics of  randomly coupled chaotic maps has been introduced.

We have found that for low connectivity in the network the transition to turbulence
via spatiotemporal intermittency in randomly coupled chaotic maps occurs as a power
law close to a critical value of the coupling.  In contrast, a discontinuous jump in
the mean turbulent fraction $\langle F \rangle$ takes place for medium and large
connectivities. Previous studies of Chat\'{e}-Manneville maps diffusively coupled on
regular lattices \cite{CM,Chate1} and on dterminsitic fractal lattices \cite{CK2} have
shown that $\langle F \rangle$ exhibits a scaling behavior close to the critical
coupling. On the other hand, $\langle F \rangle$ displays a discontinuous transition
if these maps are globally coupled \cite{CP}.

As the connectivity increases, a synchronization of the system towards the uniformly
laminar state occurs at another critical value of the coupling
$\varepsilon'_c>\varepsilon_c$. Similarly to the case of globally coupled
Chat\'e-Manneville maps \cite{CP}, windows of turbulence are established. The onset of
relaminarization appears as a discontinuous jump to a uniformly laminar state if the
connectivity is around $k=10$, and as a power law decay of the mean turbulent fraction
$\langle F \rangle$ close to $\varepsilon'_c$ for $k>10$. We have shown that the
turbulent windows contracts with increasing connectivity $k$, until they vanish for
$k\geq 100$. Additionally, periodic collective behavior arises within the windows of
turbulence, as in globally coupled maps.

Although randomly coupled maps and globally coupled maps have different topological
properties, our results show that these two classes of networks behave collectively in
analogous ways. Discontinuous phase transitions, well defined turbulent windows and
nontrivial collective behavior are common and distinctive features emerging in both
classes of networks. The recent observations of dynamical clustering in a randomly
coupled map lattice \cite{MM}, which are commonly seen is globally coupled maps, also
contribute to support the idea of the equivalence of both kinds of networks at a
global level.

The observed collective properties of the system have been analyzed through the
thermodynamic formalism. We have considered the Gibbs potential and the free energy
function for the system of randomly coupled maps and derive a closed equation for
them. Some properties of the solutions for this equation have been analyzed. In
particular, it has been proved that in some interval of the system parameters (i.e.,
coupling, connectivity, and the local parameter of the Chat\'{e}-Manneville map) the
Gibbs potential together with the free energy function acquires a two-parameter set of
branches. The non-uniqueness of the Gibbs state with respect to the given potential in
fact predicts a complex collective periodic behavior.

\section*{Acknowledgements}
\label{sec:acknow}
\noindent

The authors are grateful to the participants of seminars in  the
Zentrum f\"{u}r Interdisziplin\"{a}re Forschung, Universit\"{a}t
Bielefeld, especially to R. Lima and A. Pikovsky for illuminating
discussions. This work has been performed in connection to the
international research project "The Sciences of Complexity: From
Mathematics to technology to a Sustainable World",  Zentrum
f\"{u}r Interdisziplin\"{a}re Forschung (ZIF), Universit\"{a}t
Bielefeld. One of the authors (D.V.) has been supported by the
Alexander von Humboldt Foundation (Germany). S. Sequeira has been
supported by Graduiertenkolleg Strukturbildungsprozesse,
Forschungsschwerpunkt Mathematisierung Strukturbildungprocesse,
Universiy of Bielefeld (Germany). M.G.C. acknowledges support
from the Consejo de Desarrollo Cientifico, Humanistico y
Tecnol\'ogico at Universidad de Los Andes (Venezuela).


\section*{Figures}
\label{sec:figures}
\noindent

\newpage


\end{document}